\documentclass[fleqn,12pt]{article}
\usepackage{epsfig}
\usepackage{latexsym,amsmath,tabularx,amssymb,bm}
\usepackage{graphicx}
\usepackage{amssymb}
\usepackage{makeidx}
\usepackage{graphicx}
\usepackage{multicol}

\def\bfgrad{\mbox{\boldmath$\grad$}}

\newcommand{\nn}{\nonumber\\ }
\newcommand{\beq}{\begin{eqnarray}}
\newcommand{\eeq}{\end{eqnarray}}
\newcommand{\be}{\begin{eqnarray}}
\newcommand{\ee}{\end{eqnarray}}
\newcommand{\BQ}{\begin{equation}}
\newcommand{\EQ}{\end{equation}}
\newcommand{\BQA}{\begin{eqnarray}}
\newcommand{\EQA}{\end{eqnarray}}

\def\labe{\label}
\def\simge{\mathrel{%
   \rlap{\raise 0.511ex \hbox{$>$}}{\lower 0.511ex \hbox{$\sim$}}}}
\def\simle{\mathrel{
   \rlap{\raise 0.511ex \hbox{$<$}}{\lower 0.511ex \hbox{$\sim$}}}}
\def\bigs{\mathrel{
   \rlap{\raise 0.531ex \hbox{$>$}}{\lower 0.531ex \hbox{$<$}}}}

\def\grad{\nabla}                               
\def\del{\partial}                              

\begin{document}

\begin{titlepage}

\begin{flushright}
SACLAY--T03/124\\
CU--TP--1094
\end{flushright}

\vspace{1cm}

\begin{center}
{\LARGE\sf From  Color Glass to Color Dipoles 

\vspace{0.5cm}
in high-energy onium--onium scattering}

\vspace{0.9cm}

{\large E. Iancu$^{\rm a}$ and A. H. Mueller$^{\rm b,}$\footnote{This
research is supported in part by the US Department of Energy.} }\\

\vspace{5mm}

{\it $^{\rm a}$~Service de Physique Theorique, CE Saclay, F-91191 
        Gif-sur-Yvette Cedex, France}

\vspace{0.1cm}

{\it $^{\rm b}$ Department of Physics, Columbia University,
New York, NY 10027, USA}

\vspace{0.5cm}

\end{center}

\vspace{0.5cm}

\begin{abstract}
Within the Color Glass formalism, we construct the wavefunction
of a high energy onium in the BFKL and large--$N_c$ approximations, 
and demonstrate the equivalence with the corresponding result in the 
Color Dipole picture.
We propose a simple factorization formula for the elastic scattering 
between two non--saturated ``color glasses'' in the center--of--mass frame.
This is valid up to energies which are high enough to allow for a study 
of the onset of unitarization via multiple pomeron exchanges.
When applied to the high energy onium--onium scattering, this formula 
reduces to the Glauber--like scattering between two systems of dipoles,
in complete agreement with the dipole picture. 

\end{abstract}

\end{titlepage}
\section{Introduction}

Over the last decade, two different formalisms --- the Color Dipole Picture
(CDP) [1--12] and the Color Glass Condensate (CGC) [13--24] --- have been 
developed to study high energy scattering in QCD. Both formalisms aim at a
description of unitarization effects within perturbative QCD. But the specific
technical realizations are quite different, and so are also the 
corresponding physical pictures.
From the point of view of perturbative QCD, these are both {\it leading
logarithmic} formalisms, in the sense that they resum the radiative
corrections which are enhanced by powers of $\ln s$, with $s$ the total
invariant energy squared. But unlike the BFKL equation \cite{BFKL}, to which they both
reduce in the linear, or single scattering,
approximation (``single pomeron exchange''), 
these formalisms allow also for {\it non--linear effects}, like
multiple scattering, which are responsible for the unitarization of
the scattering amplitude. 
Still, the way how the BFKL physics and the unitarization effects 
are encoded differs substantially from one formalism to the other.
Thus, although expected, the equivalence between these two descriptions 
is by no means obvious (at least, not beyond the linear, BFKL, approximation).
It is the main purpose of this paper to demonstrate this equivalence
in the most explicit way, namely by using the Color Glass formalism to
rederive the picture of high--energy 
onium--onium scattering that has been originally obtained within
the Color Dipole formalism.

A hint towards such an equivalence comes already from the
fact that the equations for the non--linear evolution of the
scattering amplitude are rather similar, although not exactly the same,
in the two formalisms. For the CGC formalism, these are the Balitsky 
equations\footnote{These equations have been originally derived by Balitsky 
\cite{B}, within a formalism using the operator--product expansion of Wilson
line operators near the light--cone. Subsequently, Weigert has shown \cite{W}
that this infinite hierarchy of equations can be compactly summarized 
into a single {\it functional} equation. Within the CGC formalism, the 
first equation in the hierarchy by Balitsky 
has been explicitly derived in Ref. \cite{PI},
and the complete equivalence with the functional equation by Weigert
(as far as the evolution of Wilson line operators is concerned) has been
demonstrated in Ref. \cite{BIW}.} \cite{B}, 
which form an infinite hierarchy : 
With increasing energy, the original projectile ---
say, a quark--antiquark pair in a colorless state, or ``color dipole'' --- may
radiate a gluon, so the scattering amplitude for the
$q\bar q$ pair is naturally coupled in the evolution equation to the
corresponding amplitude for the $q\bar q g$ system, and so forth.
Within the CDP formalism, Kovchegov has managed to obtain a {\it closed}
equation \cite{K}, but only after making the additional assumption that
the color charges inside the target are 
uncorrelated. Kovchegov's equation is formally similar to the first equation in the
hierarchy by Balitsky, and may be viewed as an approximation to the latter,
but clearly it cannot be equivalent to it (since a closed equation
contains less dynamical information than an infinite hierarchy).
It is not a priori clear whether this lack of equivalence is intrinsic in
the two formalisms (CDP and CGC), or merely related to the
additional assumptions introduced by Kovchegov in his derivation.

Our subsequent analysis will show that the two formalisms {\it are} in fact 
equivalent, at least, for the problem of interest here
(onium--onium scattering at high energy) and within a wide (but limited) 
kinematical regime, which includes the BFKL regime and the onset of 
unitarization (see below for details). This implies that the results of
the exact, numerical, investigations of the CGC theory, which are currently 
under way \cite{RW0?}, 
should match exactly the previous Monte--Carlo simulations of the CDP, 
due to Salam \cite{Salam95}, but differ in their details \cite{IMfluct}
from the solution to the Kovchegov equation.

At this stage, it is useful to briefly discuss the two formalisms,
in order to emphasize their similitudes and
differences. In both approaches, the main ingredient is the
construction of the light--cone wavefunction of an energetic hadron in the
leading logarithmic approximation. This means that, in the wavefunction, one
keeps only the small--$x$ gluons, which form a high density system and 
multiply --- when further increasing the energy --- via the basic Lipatov vertex.

Within CDP \cite{AM94,AM95,Salam95,AMSalam96}
(see also Ref. \cite{CDPreview} for a recent review and more references), 
this construction is carried out in the large--$N_c$
limit and in the linear (or BFKL) approximation.
The large--$N_c$ limit allows one to treat a gluon like a $q\bar q$ pair in
a color octet state. Then, the emission of a gluon from a primary ``dipole''
(a $q\bar q$ pair in a colorless state) is interpreted as the original dipole 
splitting into two: each new dipole is made of the quark (antiquark) component of the
primary dipole and the antiquark (quark) component of the emitted gluon.
By iterating this elementary process, one obtains a description of the {\it evolved} 
dipole (or ``onium'') wavefunction as a system of dipoles. 
Still because of the large--$N_c$ limit, one can neglect the interference
between emissions from {\it different} dipoles: the dipoles emit gluons
independently, resulting in a tree of dipoles.
The linear approximation means that one neglects the interactions
among the emitted dipoles, so that, e.g., the dipole number density evolves
according to the linear BFKL 
equation\footnote{This is consistent with the large--$N_c$ counting,
since in the BFKL evolution
each power of $\alpha_s$ is multiplied by a factor of $N_c$.}. 
This puts a high--energy limit on the applicability of
the dipole  picture: When the dipole density, which grows 
like $N(Y) \sim {\rm e}^{\omega_0 Y}$ (with $Y \sim \ln s$ being
the rapidity, and $\omega_0 = (4\ln 2)\alpha_s N_c/\pi$), becomes so large that
$\alpha_s^2 N(Y) \sim 1$, non--linear effects like dipole recombination
become important, and are expected to lead to {\it saturation} 
\cite{GLR,MQ86,BM87,MV94,JKMW97,AM99,SAT}.

Thus, gluon saturation is not included in the CDP wavefunction, and most
likely it cannot be accomodated in this formalism (at least, not in a systematic
way), since the dipole--dipole cross--section $\sim \alpha_s^2$ is formally
of higher order in the large--$N_c$ counting. But, as we briefly recall now,
the non--linear effects responsible for unitarization {\it can} be accomodated,
namely, they can be  naturally summed up in the scattering amplitude.
This is what makes this formalism more suitable for the study of the
high--energy scattering than the conventional BFKL approach.

Specifically, consider onium--onium scattering in the center-of-mass frame 
($Y_1=Y_2 =Y/2$) at an
energy which is low enough for $\alpha_s^2 N(Y/2) \ll 1$, but high enough for
$\alpha_s^2 N^2(Y/2) \sim 1$. (Since $N(Y/2)$ is a large number, these conditions
leave a rather large window.) The first condition means that we can ignore saturation
effects in the wavefunctions of any of the incoming onia. In the second condition,
$\alpha_s^2 N^2(Y/2)$ is the probability that a pair of dipoles --- one from each
onium --- scatters with each other in the 2-gluon exchange approximation. Together
with the BFKL evolution of the individual wavefunctions, this gives the
onium--onium scattering amplitude in the ``single pomeron exchange'' (here, BFKL 
pomeron) approximation. But when $Y$ is so large that $\alpha_s^2 N^2(Y/2) \sim 1$,
the scattering is so strong that ``multiple pomeron exchanges'' (i.e., the simultaneous
scattering of several pairs of dipoles from the two onia) become equally important.
A crucial simplification, however, is that this multiple scattering refers only
to {\it different} dipoles: in the kinematical window of interest, the probability
that a {\it single} dipole undergoes multiple scattering is still suppressed, since
proportional to $\alpha_s^2 N(Y/2)$. Because of this, the multiple scattering series
can be explicitly summed up, as shown in Ref. \cite{AM95}, leading to a scattering
amplitude of a generalized Glauber type which satisfies unitarity.

Note that the judicious choice of the frame has been essential for the validity
of the previous arguments: If, instead of the center-of-mass frame, we were to choose,
say, the rest frame of the second onium ($Y_1=Y$, $Y_2 =0$), then, at rapidities large
enough for the  unitarization effects to be important, the saturation effects in
the first onium wavefunction would be important as well (since $N(Y)\simeq N^2(Y/2)$),
and the CDP formalism would not be applicable any more. We see that the distinction between
unitarization and saturation is frame dependent, and in an asymmetric frame the two
phenomena cannot be disentangled from each other.
To summarize, by working in the center-of-mass frame, CDP provides a simple description
of the onset of unitarization, while avoiding the intricacies of the non--linear
quantum evolution.
But this has the drawback that the physics of saturation cannot be studied {\it directly},
but only indirectly, via its effects on the (boost invariant) scattering amplitude.

On the other hand, the CGC formalism \cite{MV94,K96,JKMW97,JKLW97,PI,SAT}
(see also Ref. \cite{CGCreviews} for recent reviews and more references)
is precisely intended to provide a description
of the non--linear effects in the hadron wavefunction, in particular, of saturation.
In this approach, the only restriction in the construction of the wavefunction
is the leading logarithmic approximation, in the spirit of which the small--$x$ gluons
are treated as the products of radiation from fast moving ``color charges'' (the partons
with higher values of $x$), whose internal dynamics is ``frozen'' by Lorentz time
dilation (thus forming a ``color glass''). The hadron wavefunction at small--$x$ is
then fully specified by giving the probability law, or ``weight function'', for the
spatial distribution of these color charges. 

When further decreasing $x$, new quantum fluctuations become effectively frozen,
and must be included in the color source. This can be done via a perturbative QCD
calculation in which non--linear effects are taken into account via the coupling
between the quantum fluctuations and the classical color field radiated by the sources
constructed in previous steps. The result of this calculation is a functional
{\it renormalization group equation} (RGE), 
sometimes referred to as the JIMWLK equation\footnote{This stands for Jalilian--Marian, Iancu,
McLerran, Weigert, Leonidov, and Kovner, which are the authors of Refs.
\cite{JKLW97,PI,W} in which this equation has been proposed and constructed.},
which governs the evolution of the
weight function for the color sources with increasing $Y=\ln 1/x$. 

So far, only approximate solutions to this equation have been constructed,
which are valid in restricted kinematical domains\footnote{The results thus
obtained are consistent with analytic \cite{K,LT99,SCALING,AM02}
and numerical \cite{LT99,AB01,Motyka,LL01,GBS03} studies of the Kovchegov equation,
with investigations of the BFKL dynamics above $Q_s$ \cite{GLR,SCALING,MT02},
and also with previous studies of saturation \cite{GLR,JKMW97,AM99}.}. 
In the limit where the color field are weak and the non--linear effects become
negligible --- this corresponds to not so high
energies, where the gluon density is still low ---, the RGE equation has been shown
\cite{JKLW97,PI} to reproduce the BFKL equation for the gluon distribution \cite{BFKL}. 
In the opposite regime at very high energies, where the fields are strong, the
non--linear effects in the RGE 
tame the rise of the gluon distribution with $1/x$ (``saturation''), 
and lead to the formation of a high--density gluonic state --- the 
 color glass {\it condensate} --- characterized by a hard intrinsic scale, 
the saturation momentum $Q_s$, 
and by large occupation numbers, of order $1/\alpha_s$,
for all gluonic modes with momentum less than or equal to $Q_s$ \cite{SAT,GAUSS}.

One reason why these previous approximations are really crude, is that they have
been merely concerned with two--point functions (like the gluon density), while in
reality the functional RGE is equivalent to an infinite hierarchy of 
equations for the correlation functions. (This encompasses, in particular,
the hierarchy by Balitsky \cite{B,W}.) While in the strong field regime at saturation
it seems to be extremely difficult to go beyond the mean field approximation 
of Refs. \cite{SAT,GAUSS}, in the weak field regime, on the other hand,
one can rely on perturbation theory to simplify the RGE and study the coupled
evolution of the various $n$--point functions. In this perturbative regime,
one expects the RGE to reduce to the BFKL evolution, and, in particular,
to the color dipole picture at large $N_c$.

This was expected, but never proven. In this paper, we shall fill in this gap by
showing that, under the assumptions alluded to before
--- weak fields and large--$N_c$ ---, the RGE which describes the evolution of
the onium can be solved in the dipole basis, 
with a result which is indeed equivalent to that of CDP.
In fact, we shall find exactly that representation
of the onium wavefunction that has been used 
by Salam in numerical simulations of high--energy 
onium--onium scattering \cite{Salam95}. 

To emphasize that such an equivalence is not a priori obvious, let us
mention here a few technical differences between the two approaches:
Since built in terms of colorless dipoles, the CDP formalism is automatically free of
infrared singularities, but displays ultraviolet divergences (the probability to
radiate dipoles of arbitrarily small size is arbitrarily large) which cancel 
in between ``real'' and ``virtual'' contributions to physical observables.
By contrast, in the CGC formalism there are no ultraviolet divergences,
but since the corresponding degrees of freedom are now colorful, there are apparent
infrared singularities, which cancel only in the calculation of gauge invariant
quantities. By solving the RGE in the dipole basis, we shall reformulate
the CGC formalism (in the low density, or BFKL, regime)
in such a way that infrared finiteness becomes manifest.

But when comparing the two approaches, what is most interesting is the way
they describe the unitarization of high--energy scattering.
As already explained, the CDP can do that only in a {\it symmetric} frame, like
the center--of--mass frame, which for a given total energy minimizes the
importance of the saturation effects, which are not under control.
Besides, even in such a frame, CDP cannot be used at arbitrarily high energies, 
since saturation becomes eventually important. 
By contrast, the CGC formalism, which allows for non--linear effects 
in the hadron wavefunction, has no such a high energy
limitation. But its applications to scattering are generally
conditioned by the use of an {\it asymmetric} frame, 
like the infinite momentum frame of the target, in which the projectile 
has a simple structure (e.g., a $q\bar q$ pair), for
which we know how to write down the scattering amplitude.
This is why rigorous applications of this formalism have been
so far restricted to physical situations which lend themselves naturally to
such an asymmetric description, like deeply inelastic scattering, or proton--nucleus
collisions (for references see \cite{CGCreviews}). 

In principle, the fact of using different frames is not an obstacle against comparing
the two approaches: Since the scattering amplitude is boost--invariant,
this might be very well computed in an asymmetric frame within the CGC formalism,
and the result then compared to that obtained by Salam \cite{Salam95,AMSalam96}
within CDP. 
But this would require an exact calculation using the fully non--linear 
CGC wavefunction, which is not yet available, and at best could be computed 
numerically \cite{RW0?}.
To allow for an explicit, and more insightful, analytic
comparison, we find it convenient to formulate the scattering problem
in a symmetric way also within the framework of CGC.
Loosely speaking, we  shall
reformulate the CGC approach in such a way to mimic the strategy of CDP.

More specifically, we shall show that the elastic scattering between two 
{\it non--saturated} color glasses can be represented as the eikonal coupling between
the color charge in one glass and the light--cone Coulomb potential radiated by the
color charge in the other glass. (The extension of this formula to the
general case where one, or both, of the color glasses is saturated is complicated
by the fact that we do not know how to write down the coupling between an arbitrary
distribution of classical color charge and a strong non--Abelian field.)
This formulation has the same limitations,
and also the same advantages, as the CDP --- it includes multiple scattering,
but is inconsistent with non--linear effects in the hadrons wavefunctions ---,  
so it can be used too to study the onset of unitarization in 
the center--of--mass frame. The comparison between the two formalisms becomes then
straightforward, and their equivalence can be explicitly proven: When the incoming
color glasses are (non--saturated) onia, we shall
find that the CGC theory reproduces exactly 
the Glauber--like expression for the scattering amplitude originally derived by Mueller
\cite{AM95}, within the operator formulation of CDP. 

The plan of this paper is as follows: In Sect. 2, we shall construct the CGC description
of an elementary dipole, and of the dipole--dipole scattering in the two gluon exchange
approximation. The dipole wavefunction that we shall derive here will represent
the initial condition for the quantum evolution with $Y$ to be discussed in Sect. 3.
Specifically, in Sect. 3, we shall show that, in the BFKL approximation and the 
large--$N_c$ limit, the RGE for the quantum evolution of the color glass can be solved in
the dipole basis (provided the initial condition is a dipole too).
We shall thus recover the onium wavefunction of the color dipole approach.
Finally, in Sect. 4, we shall propose a factorization formula for the scattering
between two non--saturated color glasses, and show that this reproduces the 
CDP formula for onium--onium scattering in the center--of--mass frame.

\section{An elementary dipole as a color glass}
\setcounter{equation}{0}

In this section, we shall show that, for the purposes of high--energy scattering,
an {\it elementary dipole} (i.e., a quark--antiquark pair in a color single state)
can be described as a {\it color glass} (i.e., a random distribution of classical color
sources with a specific weight function). In the spirit of the eikonal approximation,
we shall assume the quark and the antiquark to be pointlike ``particles'' which propagate 
at (nearly) the speed of light, and whose transverse coordinates (with respect to the
propagation axis) are frozen: ${\bm x}$ for the quark, and ${\bm y}$ for the antiquark.
Since both the number of color charges and their spatial distribution are completely
fixed, we anticipate that the only source of randomness --- which makes the ``glassy''
description natural --- refers to the color degrees of freedom: The fact that the dipole
is a color singlet means that any average over the dipole wavefunction must include
an average over color. In more conventional calculations, this averaging is performed
by taking the trace over the color matrices which enter the coupling of the quark
or the antiquark to an ``external probe''. (We emphasize that, in a scattering problem,
this color averaging should be performed already at the amplitude level, and not only
in the cross--section.) Alternatively, as we shall see, this averaging can be formulated
as an integral over a set of random variables, with the interpretation of ``classical color
charges''. In what follows, we shall be mainly interested in the lowest--order
scattering processes, which proceed via two gluon exchange; in that case, the random
variables can be taken as Gaussian.

To remain as simple as possible, we shall consider the scattering between two dipoles
in the two--gluon exchange approximation, for which the result is well known.
Here, we shall rephrase this standard result as the collision between two 
color glasses, in a form which is suitable for further generalizations,
like the inclusion of quantum evolution and multiple scattering.

To start with, we shall consider a more general process,
whose description is well established in the
CGC formalism, and which encompasses the dipole--dipole
scattering as a special case, as we shall shortly see: 
This is the scattering between an elementary dipole
and a color glass. The following discussion will also give us the opportunity 
to recall the basic ingredients of the CGC approach, and fix some notations.

Quite generally, the $S$--matrix element for a head-on dipole--hadron 
collision can be computed in the eikonal approximation as:
\be\label{Sdef}
S({\bm{x}},{\bm{y}})\,\equiv\,\frac{1}{N_c}\,
\Big\langle {\rm tr}\big(V^\dagger({\bm{x}}) V({\bm{y}})\big)
\Big\rangle,\ee
where $V^\dagger({\bm{x}})$ and $V({\bm{y}})$ are Wilson lines
describing the scattering of the quark, or the antiquark, off the color
field in the hadron, and the color trace divided by $N_c$ is the average over 
color alluded to before. Furthermore, the brackets in the right hand side indicate 
the average over the hadron wavefunction. The CGC theory provides an explicit realization
for this average, namely (see, e.g., \cite{CGCreviews} for more details) :
\be\label{CGCaverage}
\frac{1}{N_c}\,
\Big\langle {\rm tr}\big(V^\dagger({\bm{x}}) V({\bm{y}})\big)\Big\rangle\,=\,
\int\,{\rm D}[\alpha]\, \,W[\alpha]\,\,\frac{1}{N_c}\,
{\rm tr}\Big(V^\dagger_{{\bm{x}}}[\alpha]\, V_{{\bm{y}}}[\alpha]\Big),
\ee
where $\alpha^a(x^-,{\bm{x}})$ is the light--cone Coulomb field radiated by the
color sources in the hadron, $W[\alpha]$ is a positive-definite functional which
specifies the probability to find a given field configuration 
(the ``weight function''), and:
\begin{equation}
\label{Vdef}
V^\dagger_{\bm{x}}[\alpha]
\,\equiv\,{\rm P}\,{\rm exp}\left({\rm i}g\int dx^-
\alpha^a(x^-,{\bm{x}}) t^a\right). 
\end{equation}
with P denoting path--ordering in $x^-$. Note that we are using light--cone vector
notations, $x^\pm \equiv (t\pm z)/\sqrt{2}$, and our conventions are such that the
hadron is a right mover (it propagates in the positive $z$, or positive $x^+$, 
direction), while the dipole is a left mover (negative $z$, or positive $x^-$).

The expression (\ref{CGCaverage}) is written in a specific gauge, namely the
covariant gauge, in which $\alpha^a$ is the only non--zero component of the field
in the hadron ($A^\mu_a = \delta^{\mu +}\alpha^a$), and is time--independent (i.e., 
independent of $x^+$). The relation between this field and the color charge density
in the hadron  $\rho^a(x^-,{\bm{x}})$ is simply given by the 
two--dimensional Poisson equation:
\be\label{Poisson}
-\grad_\perp^2 \alpha^a(x^-,{\bm{x}})\,=\,\rho^a(x^-,{\bm{x}}).\ee
But the quantity computed in Eq.~(\ref{CGCaverage}) is actually
gauge--invariant, since so are both the weight function $W[\alpha]$ and the scattering
operator built from Wilson lines \cite{CGCreviews}.

Eq.~(\ref{CGCaverage}) holds in any frame in which the dipole rapidity is not
too large, so that one can neglect gluon radiation in the dipole wavefunction.
If $Y$ is the total rapidity gap, with $Y=y_{\rm hadron} + |y_{\rm dipole}|$, the precise
condition reads (recall that we are working in a leading logarithmic approximation) :
$\alpha_s |y_{\rm dipole}|\ll 1$. On the other hand, there is no restriction on the rapidity
$y_{\rm hadron}$ of the hadron. If $\alpha_s y_{\rm hadron} \simge 1$, this means that
the effects of the quantum evolution must be included in the weight function, which
to this purpose must be a function of $y_{\rm hadron} \approx Y$. The evolution of the
weight function $W[\alpha]\equiv W_Y[\alpha]$ with $Y$ will be discussed in the next section.
Here, we are only interested in the simple situation in which the hadronic target
(i.e., the ``glass'' in Eq.~(\ref{CGCaverage})) is itself an elementary dipole.
This entails several simplifications:

First, the color field of a dipole is {\it weak\,}. This means that the fluctuating
field $\alpha$ in Eq.~(\ref{CGCaverage}) has typically small amplitudes,
$g\alpha \ll 1$, so we can expand the Wilson lines in powers of $g\alpha$. 
The lowest non--trivial contribution to $S$ is of order\footnote{We refer here to the
powers of $g$ which are explicit in the expansion of the Wilson lines for fixed $\alpha$
(with $g\alpha \ll 1$). In the final result for $S$, additional factors of $g$ may arise
from the evaluation of the correlation functions of $\alpha$ as in Eq.~(\ref{CGCaverage}).}
$g^2$, and is obtained by expanding  the Wilson lines to quadratic order in $g\alpha$.
(The terms of order $g$, which would be linear in $\alpha^a t^a$,
vanish after averaging with the gauge--invariant weight function, or,
alternatively, after taking the color trace.) We have:
\be
\label{Vexp}
V^\dagger_{\bm{x}}[\alpha]
&\approx &1\,+\, {\rm i}g\int dx^-
\alpha^a(x^-,{\bm{x}}) t^a\\
&-&\frac{g^2}{2} \int dx^-\!\int dy^-
\alpha^a(x^-,{\bm{x}})\alpha^b(y^-,{\bm{x}})\big[\theta(x^- - y^-)t^a t^b
+ \theta(y^- - x^-)t^b t^a\big].\nonumber \ee
In forming $S$, Eq.~(\ref{Sdef}), we see that the {\it ordering} of the color matrices
in $x^-$ becomes irrelevant in the present approximation, because of the symmetry
of the color trace: 
$${\rm tr} (t^a t^b)\,=\,\frac{1}{2}\,\delta^{ab}\,=\,{\rm tr} (t^b t^a)\,.$$
We thus obtain:
\be\label{Sexp}
S({\bm{x}},{\bm{y}})\,\approx\,1 - \,\frac{g^2}{4N_c}\,\Big\langle \big(\alpha^a({\bm{x}})
- \alpha^a({\bm{y}})\big)^2\Big\rangle\,+\,{\cal O}(g^3),\ee
where:
\be\label{alphaT}
\alpha^a({\bm{x}})\,\equiv\,\int dx^-\,\alpha^a(x^-,{\bm{x}})\ee
is the effective color field in the transverse plane, as obtained after integrating
over the longitudinal profile of the hadron. We thus see that, in this approximation,
the {\it longitudinal structure} of the color field becomes irrelevant as well (this is,
of course, correlated with the fact that the  ordering in $x^-$ is unimportant).
To evaluate the average in Eq.~(\ref{Sexp}), it is therefore enough
to consider the {\it reduced} weight function which specifies the distribution of the
field $\alpha^a({\bm{x}})$ in the transverse plane alone.

Eq.~(\ref{Sexp}) 
holds whenever the color fields in the target are weak.
In this regime, and to lowest order in perturbation theory, the scattering amplitude
$T = 1-S$ is determined solely by the two--point function of the color fields,
which in turn can be traded in the (unintegrated) gluon distribution.
In particular, in the case where the target is itself a dipole, Eq.~(\ref{Sexp})
must reproduce the standard result for the dipole--dipole
scattering in the two--gluon exchange approximation. This condition
constrains the weight function $W[\alpha]$ which describes a dipole
as a color glass, and can be used to actually construct 
this weight function (which, in the present approximations, is simply a Gaussian).

Here, we shall perform this construction via a slight detour, which will allow us
to introduce an intuitive representation for the color charge of a dipole.
With this aim, we shall take advantage of the fact that the final expression for $S$ must
be symmetric in the two dipoles. So, our first goal will be to rewrite Eq.~(\ref{Sexp})
in such a way to make this symmetry manifest. Note that, to the order of interest, 
Eq.~(\ref{Sexp}) is equivalent to:
\be\label{Seikonal0}
S({\bm{x}},{\bm{y}})\,\approx\,\frac{1}{N_c}\,\Big\langle{\rm tr}\,{\rm e}^{\,{\rm i}g
(\alpha^a({\bm{x}})- \alpha^a({\bm{y}}))t^a}\Big\rangle \,=\,
\frac{1}{N_c}\,\Big\langle{\rm tr}\,{\rm e}^{\,{\rm i}\int d^2{\bm z}\,\hat\rho^a_L({\bm z})
\alpha^a({\bm{z}})}\Big\rangle,\ee 
which is formally the same as Eq.~(\ref{Sdef}) but without the path-ordering.
This rewriting has naturally introduced the following expression for
the color charge density of the left--moving dipole:
\be\label{rhoL}
\hat\rho^a_L({\bm z})\,\equiv\,gt^a\big[\delta^{(2)}({\bm z}-{\bm{x}})
- \delta^{(2)}({\bm z}-{\bm{y}})\big].\ee
The hat on $\hat\rho^a_L$ is to remind that this is a color matrix,
as opposed to the $c$--number charge density $\rho^a$ in the CGC formalism.

Clearly, a similar color charge density should be associated also with the
right--moving dipole, with the quark at ${\bm{x}_0}$ and the antiquark at ${\bm{y}_0}$ :
\be\label{hatrhoR}
\hat\rho^a_R({\bm z})\,\equiv\,gt^a\big[\delta^{(2)}({\bm z}-{\bm{x}_0})
- \delta^{(2)}({\bm z}-{\bm{y}_0})\big].\ee
This dipole generates the matrix--valued field (cf. Eq.~(\ref{Poisson})) :
\be\label{hatalpha}
\hat\alpha^a_R({\bm{z}})\,=\,gt^a\big[\Delta({\bm z}-{\bm{x}_0})
- \Delta({\bm z}-{\bm{y}_0})\big]\,\equiv\, gt^a\,{\cal G}({\bm z}|{\bm x}_0, {\bm{y}_0}),\ee
where $\Delta({\bm x}-{\bm{y}})$ is the two--dimensional Coulomb propagator:
\be\label{Delta}
\Delta({\bm x}-{\bm{y}}) \,=\,\int\! \frac{d^2{\bm k}}{(2\pi)^3}\,\frac
{{\rm e}^{\,{\rm i} {\bm k}\cdot ({\bm x}-{\bm{y}})}}{{\bm k}^2}\,=\,\frac
{1}{4\pi}\,\ln \frac{1}{({\bm x}-{\bm{y}})^2\mu^2}\,.\ee
The infrared regulator $\mu$ is needed to write down the propagator, but is harmless
in the present context, as it cancels out in the difference 
${\cal G}$ of the two propagators in  Eq.~(\ref{hatalpha}). This is a typical
infrared cancellation permitted by the colorless nature of the dipole.

These considerations suggest that, 
for dipole--dipole scattering, Eq.~(\ref{Seikonal0}) should be equivalent to
the following, manifestly symmetric\footnote{To make this symmetry even more obvious, note
that $\int d^2{\bm z}\,\hat\rho^a_L({\bm z})
\hat\alpha^a_R({\bm{z}}) = \int d^2{\bm z}\,\grad^i \hat\alpha^a_L({\bm{z}})
\grad^i\hat\alpha^a_R({\bm{z}})$.}, formula :
\be\label{Seikonal1}
S({\bm{x}},{\bm{y}}|{\bm x}_0,{\bm{y}}_0)
\,\approx\,\frac{1}{N_c^2}\,{\rm tr}_L\otimes {\rm tr}_R
\,{\rm e}^{\,{\rm i}\int d^2{\bm z}\,\hat\rho^a_L({\bm z})
\hat\alpha^a_R({\bm{z}})}.\ee 
As compared to Eq.~(\ref{Seikonal0}), the field $\alpha$ has now the explicit
expression (\ref{hatalpha}), and the average over $\alpha$ reduces to the
color trace $(1/N_c){\rm tr}_R$. (In  Eq.~(\ref{Seikonal1}),
the color traces act separately in the color spaces of the first dipole and
the second dipoles, respectively. To make that clear, one could use different notations,
say, $t^a_L$ and $t^a_R$, for the color matrices spanning these two spaces.)

Let us check that Eq.~(\ref{Seikonal1}) reproduces indeed the expected result:
After expanding this expression to lowest non--trivial order, 
i.e., to quadratic order in each of the two color charge
densities $\hat\rho^a_L$ and $\hat\rho^a_R$, and performing the color averages,
one obtains:
\be\label{Seikonal2}
S({\bm{x}},{\bm{y}}|{\bm x}_0,{\bm{y}}_0)
\,\approx\,
1 - \,\frac{g^4}{8N_c^2}\,(N_c^2-1)\,
[{\cal D}({\bm{x}},{\bm{y}}|{\bm x}_0,{\bm{y}}_0)]^2,\ee
where
\be\label{calD}
{\cal D}({\bm{x}},{\bm{y}}|{\bm x}_0,{\bm{y}}_0)&\equiv&
\Delta({\bm x}-{\bm{x}}_0) - \Delta({\bm x}-{\bm{y}}_0) - \Delta({\bm y}-{\bm{x}}_0)
+ \Delta({\bm y}-{\bm{y}}_0)\nn
&=&\frac{1}{4\pi}\,\ln \frac{({\bm x}-{\bm{y}}_0)^2\,({\bm y}-{\bm{x}}_0)^2
}{({\bm x}-{\bm{x}}_0)^2\,({\bm y}-{\bm{y}}_0)^2}\,.\ee
We recognize in Eqs.~(\ref{Seikonal2})--(\ref{calD})
the standard result for the dipole--dipole
scattering amplitude in the two--gluon exchange approximation (see, e.g., \cite{NW97}).

Eq.~(\ref{Seikonal1}) makes explicit the fact that, for a dipole, the ``average over
the hadron wavefunction'' reduces to an average over color, here implemented as a color
trace (separately for each dipole). In what follows, we shall rephrase this in the
color glass formalism. That is, we shall return to $c$--number densities and fields, and
replace the color traces with functional averages over $\alpha_L^a$ and  
$\alpha_R^a$, like in Eq.~(\ref{CGCaverage}). (It is now understood that, in a more
symmetric notation, the field variable $\alpha^a$ in Eq.~(\ref{CGCaverage}) should
be renoted as $\alpha_R^a$.)

We thus introduce  $c$--number random color charge distributions of the form:
\be\label{rhoR}
\rho^a_R({\bm z})\,\equiv\,Q_R^a\big[\delta^{(2)}({\bm z}-{\bm{x}}_0)
- \delta^{(2)}({\bm z}-{\bm{y}}_0)\big],\ee
and similarly for $\rho_L^a$, where the ``classical color charges'' $Q_R^a$ and $Q_L^a$
are Gaussian random variables with the following correlation functions ($s=L,R$) :
\be\label{QLR}
\langle Q^a_s \rangle_Q \,=\,0,\qquad \langle Q^a_s Q^b_{s'}\rangle_Q \,=\,
\delta^{ab}\delta_{ss'} \lambda,\qquad \lambda\,\equiv\,\frac{g^2}{2N_c}\,.\ee
Note that $Q^a$ corresponds to $gt^a$, and the two--point function in Eq.~(\ref{QLR})
has been chosen to match $(1/N_c) {\rm tr} (t^at^b)$. It is easy to check that, 
to order $g^4$, Eq.~(\ref{Seikonal1}) is equivalent to
\be\label{Seikonal3}
S({\bm{x}},{\bm{y}}|{\bm x}_0,{\bm{y}}_0)
\,=\,\Big\langle\,{\rm e}^{\,{\rm i}\int d^2{\bm z}\,\rho^a_L({\bm z})
\alpha^a_R({\bm{z}})}\Big\rangle_Q \,\ee
with (cf. Eq.~(\ref{Poisson})) :
\be\label{alphaR}
\alpha^a_R({\bm{z}})\,=\,Q_R^a\big[\Delta({\bm z}-{\bm{x}}_0)
- \Delta({\bm z}-{\bm{y}}_0)\big]\,\equiv\,Q_R^a\, {\cal G}({\bm z}|{\bm{x}}_0,{\bm{y}}_0).\ee

Then, we replace the $Q$--average in Eq.~(\ref{Seikonal3}) by 
functional averages like in Eq.~(\ref{CGCaverage}). Specifically, Eq.~(\ref{Seikonal3}) 
is the same as:
\be\label{CGCDD}
S({\bm{x}},{\bm{y}}|{\bm x}_0,{\bm{y}}_0)
\,=\,\int {\rm D}[\alpha_R]\, \,W_0[\alpha_R]\int {\rm D}[\alpha_L]\, \,W_0[\alpha_L]
\,\,{\rm e}^{\,{\rm i}\int d^2{\bm z}\,\grad^i \alpha^a_L({\bm{z}})
\grad^i\alpha^a_R({\bm{z}})}\,,
\ee
with the following color glass weight function for an elementary dipole:
\be\label{DWF}
W_0[\alpha]\,\equiv\,{\cal N}\int \prod_{a=1}^{N_g}dQ^a\,\,{\rm exp}\left\{-\frac{Q^aQ^a}{2\lambda}
\right\}
\,\delta\big[\alpha^a({\bm{z}}) - Q^a {\cal G}({\bm z}|{\bm{x}}_0,{\bm{y}}_0)\big]\ee
In Eq.~(\ref{DWF}), $N_g=N_c^2-1$,
${\cal N}$ is a normalization factor, chosen such that
$\int {\rm D}[\alpha]\, W_0[\alpha] = 1$, and the functional $\delta$--function is
understood with a discretization of the transverse plane:
$$ \delta\big[\alpha^a({\bm{z}}) - Q^a {\cal G}({\bm z}|{\bm{x}}_0,{\bm{y}}_0)\big]
\,=\,\prod_{{\bm{z}}} \,\prod_{a=1}^{N_g}\,\delta
\big(\alpha^a({\bm{z}}) - Q^a {\cal G}({\bm z}|{\bm{x}}_0,{\bm{y}}_0)\big) $$
Clearly, the weight function $W_0[\alpha]$ depends also upon the dipole
transverse coordinates
${\bm{x}}_0$ and ${\bm{y}}_0$, but this dependence is suppressed in its notation, for
simplicity.

The interaction piece ${\rm exp}\{{\rm i}\int \!d^2{\bm z}\,\rho^a_L({\bm z})
\alpha^a_R({\bm{z}})\}$ in the previous formulae,
Eqs.~(\ref{CGCDD}) or (\ref{Seikonal3}), may be recognized as
the eikonal coupling between the color charge density in one system and the field
created by the color charge of the other system. In the Abelian case
(i.e., for electromagnetic dipoles), this would be the exact coupling at high energy,
{\it including multiple collisions}. But in the non--Abelian case, this is correct
only to order $g^4$ (corresponding to a single scattering), as derived above.
This is already clear from the fact that, in QCD, this interaction term is not
gauge invariant. In general, to describe multiple (eikonal) scattering in QCD
one has to use path--ordered exponentials, like in Eqs.~(\ref{Sdef})--(\ref{Vdef}),
to account for the non--commutativity of the color matrices in the interaction vertices. 
Still, in Sect. 4 below we shall argue that a simple factorized formula like
Eq.~(\ref{CGCDD}) can be used also for multiple scattering, but
only in a symmetric frame and within a limited range of energies.

To conclude this section, let us derive an alternative expression for the 
dipole weight function, Eq.~(\ref{DWF}), which will be more useful for what follows.
Namely, since the color field $Q^a {\cal G}$ of a dipole is weak, of order $g$
(cf. Eqs.~(\ref{QLR}) and (\ref{alphaR})), it is possible to expand the $\delta$--functional 
in Eq.~(\ref{DWF}) to quadratic order in $Q^a {\cal G}$ without loss of accuracy.
Then, one can explicitly perform the average over the classical color charges, and
thus deduce the following formula, which is our final result in this section:
\be\label{DWFexp}\hspace*{-.5cm}
W_0[\alpha]\,=\,{\cal N}\left\{1 + \frac{g^2}{4N_c}
\int \! d^2{\bm u} \!\int \! d^2{\bm v}\, \,{\cal G}({\bm u}|{\bm{x}}_0,{\bm{y}}_0)
{\cal G}({\bm v}|{\bm{x}}_0,{\bm{y}}_0)\,\frac{\delta^2}{\delta \alpha^a({\bm{u}})
\delta \alpha^a({\bm{v}})}\right\}\delta[\alpha].\ee

\section{BFKL evolution: From color glass to color dipoles} 
\setcounter{equation}{0}

The main advantage of using a color glass description for an elementary dipole
is that one can rely on the whole machinery of the CGC formalism 
\cite{JKLW97,PI,SAT,AM01} to study the
evolution of the dipole wavefunction with increasing energy (or rapidity $Y$).
As mentioned in the Introduction, this amounts to solving a functional renormalization
group equation (RGE) for the weight function $W_Y[\alpha]$ with the initial condition
$W_0[\alpha]$ given by Eq.~(\ref{DWF}) or (\ref{DWFexp}). In principle, this equation
can be used to compute $W_Y[\alpha]$ up to arbitrarily high $Y$, including in the
non--linear regime at saturation. In practice, however, the complicated 
structure of the general RGE prevents us from obtaining explicit solutions, except
under very crude approximations \cite{SAT,GAUSS}, or through numerical simulations 
(which present their own difficulties, though) \cite{RW0?}. So far, the solution to
the RGE has not been fully investigated not even in the weak field regime at not
so high energies, where saturation effects are unimportant, and BFKL physics
should apply. In particular, this is the regime in which it makes sense to compare
the predictions of the RGE with the Color Dipole picture of
Refs. \cite{AM94,AM95,Salam95,AMSalam96}. Although in this regime the
RGE simplifies considerably \cite{PI} (see also below), 
it remains a non--linear equation, which is still difficult to solve. 

Throughout this paper, we shall restrict ourselves to this weak field regime,
that we shall further simplify by taking the large--$N_c$ limit, in the spirit
of the dipole picture. Under these assumptions, we shall be able to construct
an explicit representation for the solution $W_Y[\alpha]$ in terms of a system of
dipoles which undergoes BFKL evolution. In this representation, the equivalence 
with the color dipole picture of the onium wavefunction will become transparent.

Specifically, we shall find that, under the weak field and large--$N_c$ approximations,
the RGE can be solved with the following Ansatz (compare to Eq.~(\ref{DWF})):
\be\label{WFY}
W_Y[\alpha]&=&\sum_{N=1}^\infty\int d\Gamma_N\,\,P_N(x_1,y_1;x_2,y_2;\dots;x_N,y_N|Y)
\nn &{}&\times
\int \prod_{i=1}^N \prod_{a=1}^{N_g}dQ^a_i\,\,{\rm e}^{-\frac{Q^a_iQ^a_i}{2\lambda}}
\,\delta\left[\alpha^a({\bm{z}}) - \sum_{i=1}^N Q^a_i 
{\cal G}({\bm z}|{\bm{x}}_i,{\bm{y}}_i)\right],\ee
where the notations are as follows:
$P_N(x_1,y_1;x_2,y_2;\dots;x_N,y_N|Y)$ is the probability density to generate a system
of $N$ dipoles with given transverse coordinates (namely, $({\bm{x}}_i,{\bm{y}}_i)$ for the
$i$th dipole) via the quantum evolution of an original dipole with coordinates
$({\bm{x}}_0,{\bm{y}}_0)$ through a rapidity interval equal to $Y$. (Of course, $P_N$
depends also upon the original  coordinates $({\bm{x}}_0,{\bm{y}}_0)$, and so does 
$W_Y[\alpha]$, but this dependence is kept implicit, to simplify the notation.)
The dipoles are characterized also by the respective color charges ($Q^a_i$
for the quark and $-Q^a_i$ for the antiquark, like in Eq.~(\ref{hatrhoR})), but their
distribution in color factorizes from that in the transverse space, and is separately
a Gaussian for each dipole. 
The $\delta$--functional enforces the total field in the system to be precisely 
the field generated by the $N$ dipoles (in a given color configuration). Finally,
there is a sum over configurations, which includes the average over color 
(performed separately for each dipole), the integral $\int d\Gamma_N$
over the tranverse coordinates of the $N$--dipole system, and the sum over $N$.
The probabilities $P_N(Y)$ are determined by solving a linear system of coupled
evolution equations which follow from the functional RGE after inserting the Ansatz
(\ref{WFY}) for the solution. As we shall see, these are precisely the equations 
used by Salam \cite{Salam95}
in his numerical construction of the onium wavefunction based on the CDP.

\subsection{The RGE in the BFKL approximation}

In this subsection, after briefly recalling the general structure of the RGE,
we shall derive its weak field, or BFKL, approximation, and discuss the
simplifications which occur in this limit.

The RGE is a functional Fokker--Planck equation, that is, a second--order 
functional differential equation of the diffusion type. It reads \cite{PI}
(see also Ref. \cite{JKLW97} for an early version of this equation, and Refs. \cite{W,AM01} 
for alternative derivations using the approach pioneered by Balitsky \cite{B}) :
\be\labe{RGE}
{\del W_Y[\alpha] \over {\del Y}}\,=\,{1 \over 2}\int_{{\bm{x}},{\bm{y}}}\,
{\delta \over {\delta \alpha_Y^a({\bm{x}})}}\,
\eta^{ab}({\bm{x}},{\bm{y}})[\alpha] \,{\delta W_Y\over 
{\delta \alpha_Y^b({\bm{y}})}}\,,\ee
where $\int_{\bm{x}} \equiv \int d^2 {\bm x}$, and
 the kernel $\eta^{ab}({\bm{x}},{\bm{y}})$ is a positive--definite and non--linear
functional of $\alpha$, upon which it depends via Wilson lines:
\be\label{eta}
\eta^{a b}({\bm{x}},{\bm{y}})[\alpha] \,=\,\int\!\frac{d^2 {\bm z}}{\pi}\,
 {\cal K}({\bm{x}, \bm{y}, \bm{z}})
\,(1- \tilde V^\dagger_{\bm{z}}  \tilde V_{\bm{x}})^{fa}(1 -  \tilde V^\dagger_{\bm{z}}
  \tilde V_{\bm{y}})^{fb}.
   \ee
In this equation,
\be\label{calK}
{\cal K}({\bm{x}, \bm{y}, \bm{z}}) \equiv \frac{1}{(2\pi)^2}\,
   \frac{(\bm{x}-\bm{z})\cdot(\bm{y}-\bm{z})}{
     (\bm{x}-\bm{z})^2 (\bm{z}-\bm{y})^2}\,=\,\grad^i_{\bm{z}}\Delta({\bm{x}}-{\bm{z}})
\grad^i_{\bm{z}}\Delta({\bm{y}}-{\bm{z}}),\ee
and $\tilde V^\dagger$ and $\tilde V$ are Wilson lines in the {\it adjoint} representation,
given by Eq.~(\ref{Vdef}) with $t^a\to T^a$.
It is worth recalling here that the right hand side of Eq.~(\ref{RGE}) has been obtained
after combining ``real'' and ``virtual'' contributions to the quantum evolution: The ``real''
contribution is represented by $\eta$, while the ``virtual'' one is generated as the
functional derivative of $\eta$ with respect to $\alpha$.

Note the label $Y$ on the field argument $\alpha_Y$ of the functional
derivatives in  Eq.~(\ref{RGE}): As explained in \cite{PI,CGCreviews}, this
specifies the longitudinal coordinate $x^-$ at which the functional
derivatives are to be taken. 
However, this prescription becomes irrelevant in the weak field regime of interest,
in which the (lowest order) dynamics is sensitive only to the projection of the
field in the transverse plane, as defined in Eq.~(\ref{alphaT}).
Indeed, when $g\alpha \ll 1$, one can expand the Wilson lines
in Eq.~(\ref{eta}) in perturbation theory, and obtain, e.g.,
\be\label{expV}
(1- \tilde V^\dagger_{\bm{z}}  \tilde V_{\bm{x}})^{fa}\,\approx\,ig\big(\alpha^c({\bm{x}})
- \alpha^c({\bm{z}})\big) \big(T^c \big)_{fa},\ee
which yields the lowest--order perturbative approximation to $\eta$,
of order $g^2$ :
\be\label{etalin}
\eta^{a b}({\bm{x}},{\bm{y}}) \,\simeq\,g^2\big(T^c T^d\big)_{ab}
\int\!\frac{d^2 {\bm z}}{\pi}\, {\cal K}({\bm{x}, \bm{y}, \bm{z}})\,\big[
\alpha^c({\bm{x}}) - \alpha^c({\bm{z}})\big]\big[
\alpha^d({\bm{y}}) - \alpha^d({\bm{z}})\big].\ee
As anticipated, Eqs.~(\ref{expV})-(\ref{etalin}) are insensitive to the longitudinal structure of
the field, and so are also the observables computed in this approximation,
like the dipole scattering amplitude (\ref{Sexp}). This reinforces our conclusion
in Sect. 2 that, in the weak field regime, it is enough to work with the {\it reduced}
weight function which is a functional of $\alpha^a({\bm{x}})$ alone.
This is the functional that we shall denote as
$W_Y[\alpha]$ in what follows. Correspondingly, 
the argument of any functional derivative 
will be interpreted as $\alpha^a({\bm{x}})$. 

The disappearance of the longitudinal coordinate from the problem is the first 
simplification specific to the weak field limit. The second simplification
is that the kernel $\eta$ becomes just {\it quadratic}
in $\alpha$, as manifest on Eq.~(\ref{etalin}). Clearly, even with this kernel, the RGE
(\ref{RGE}) remains non--linear, but the non--linearity is now  considerably simpler
than with the general kernel (\ref{eta}). This is best appreciated by inspection of
the evolution equations satisfied by the $n$--point functions 
$\langle \alpha(1)\alpha(2)\cdots \alpha(n)\rangle_Y$, which are obtained as follows:
Start with the general definition of a correlation function in the CGC formalism:
\be\label{OBS}   \langle {\cal O}\rangle_Y\,=\,
\int\,{\rm D}[\alpha]\, \,W_Y[\alpha]\,\,{\cal O}[\alpha],\ee
with ${\cal O}[\alpha] = \alpha(1)\alpha(2)\cdots \alpha(n)$, then take a
derivative with respect to $Y$, use Eq.~(\ref{RGE}) for $\del W_Y/\del Y$, and integrate
twice by parts in the functional integral, to finally obtain:
\be\label{Oevol}
\frac{\del \langle {\cal O}\rangle_Y}{\del Y}\,=\,
\left\langle{1 \over 2}\int_{{\bm{x}},{\bm{y}}}\,
{\delta \over {\delta \alpha^a({\bm{x}})}}\,
\eta^{ab}({\bm{x}},{\bm{y}}) \,{\delta \over 
{\delta \alpha^b({\bm{y}})}}\,{\cal O}[\alpha]
\right\rangle_Y\,,\ee

If $\eta$ is the general kernel (\ref{eta}), the r.h.s. of Eq.~(\ref{Oevol}) involves
$n$--point functions with arbitrary $n$, as generated by the expansion of the Wilson lines.
Thus, the general RGE is equivalent to an intricate hierarchy of coupled evolution 
equations\footnote{This hierarchy becomes somehow simpler if the equations are written for the
correlation functions of the Wilson lines (rather than $\alpha$). The resulting 
equations are those originally derived by Balitsky \cite{B}.},
which are generally non--linear and must be solved simultaneously (since, e.g., the evolution of
the 2--point function is coupled to that of all the $n$--point functions with $n\ge 2$).
By contrast, with the quadratic kernel in Eq.~(\ref{etalin}), the evolution equation for
correlation functions are {\it linear}, and do not mix  $n$--point functions with 
different number of fields $n$.  So far, only the equation satisfied by the 2--point 
function has been considered in the literature \cite{JKLW97,PI,GAUSS}, and shown to be
equivalent to the BFKL equation. In what follows, we shall analyze directly the functional RGE
with kernel (\ref{etalin}) --- thus encompassing all the $n$--point correlations ---
and use the large--$N_c$ 
limit\footnote{Note that the large--$N_c$ approximation cannot be implemented at the level
of the kernel $\eta$ alone, but requires some detailed information about the color
structure of the weight function.}
to construct an explicit solution.
This is the solution anticipated in Eq.~(\ref{WFY}).

Before we conclude this subsection, there is one more issue which needs to be clarified:
the convergence of the integral
over ${\bm{z}}$ which enters the kernel $\eta$, cf. Eqs.~(\ref{eta}) or (\ref{etalin}).
Clearly, there are no short--distance singularities:
the poles in ${\cal K}({\bm{x}, \bm{y}, \bm{z}}) $ at $\bm{z} = \bm{x}$ or
$\bm{z} = \bm{y}$ are compensated by the field--dependent factors in these equations,
which vanish linearly at these points. But at large distances $z=|\bm{z}| \gg x,y$,
we have ${\cal K} \sim 1/z^2$, which is not enough to guarantee
the absence of long--range (or ``infrared'') singularities. Whether such singularities
appear or not, depends also upon the nature of the operator ${\cal O}[\alpha]$,
and upon the behavior of the $n$--point functions  of $\alpha$ at large transverse separations
(i.e., upon the properties of the weight function).

For a generic weight function, infrared finiteness has been verified so far (on specific
examples) only for gauge--invariant operators, like the $S$--matrix element in Eqs.~(\ref{Sdef})
or (\ref{Sexp}). However, we shall see below that, for the onium weight function in 
Eq.~(\ref{WFY}), infrared finiteness is automatically ensured for {\it any} operator
(whether gauge--invariant or not), because of the rapid decay of the color field of a dipole.

\subsection{Quantum evolution: the first step}

Before attacking the full RGE at arbitrary rapidity $Y$,
let us study the very first step in the quantum evolution of a dipole.
That is, start with an elementary dipole with transverse coordinates 
$({\bm{x}}_0,{\bm{y}}_0)$ at $Y_0=0$ and study its evolution under a
rapidity increment ${\rm d}Y$, with $\alpha_s {\rm d}Y\ll 1$. Our aim 
is to show that this evolution can be viewed as the splitting of the original
dipole into two new dipoles. This elementary example will also allow us to
introduce in a simple setting some of the technical manipulations that will
be useful later, in the general case.

Since the weight function (\ref{DWF}) of an elementary dipole is characterized by
a single non--trivial correlation function, namely, the 2--point function:
\be\label{2point0}
\big\langle \alpha^a({\bm{x}}) \alpha^a({\bm{y}})\big\rangle_0 \,=\,g^2 C_F\,
{\cal G}({\bm x}|{\bm{x}}_0,{\bm{y}}_0){\cal G}({\bm y}|{\bm{x}}_0,{\bm{y}}_0),\ee
(we have also used $C_F= N_g/2N_c$), it suffices to compute the change in this quantity
in the first step of the evolution, or, equivalently, its derivative at $ Y=0$.
This is obtained by letting $Y\to 0$ in the evolution equation 
obtained by replacing ${\cal O}[\alpha]\to \alpha^a({\bm{x}}) \alpha^a({\bm{y}})$
in Eq.~(\ref{Oevol}) with kernel (\ref{etalin}). Simple algebra yields\footnote{Note
incidentally that, if Eq.~(\ref{2pointEVOL}) is used to deduce the evolution
equation for the scattering amplitude $T=1-S$ in Eq.~(\ref{Sexp}), one finds 
the BFKL equation, as it should.} :
\be\label{2pointEVOL}
\frac{{\rm d}}{{\rm d}Y}\,\big\langle \alpha^a_{\bm{x}} \alpha^a_{\bm{y}}\big\rangle_Y
\!\!&=&\!\!\frac{g^2N_c}{2\pi}\int_{\bm z}\Big\langle
2{\cal K}_{\bm{x} \bm{y} \bm{z}}\big(\alpha^a_{\bm{x}} - \alpha^a_{\bm{z}}\big)
(\alpha^a_{\bm{y}} - \alpha^a_{\bm{z}}\big) \nn 
&{}& \qquad\qquad- \,{\cal K}_{\bm{x} \bm{x} \bm{z}}
\alpha^a_{\bm{y}}\big(\alpha^a_{\bm{x}} - \alpha^a_{\bm{z}}\big)
- {\cal K}_{\bm{y} \bm{y} \bm{z}}\alpha^a_{\bm{x}}(\alpha^a_{\bm{y}} - \alpha^a_{\bm{z}}\big)
\Big\rangle_Y,\ee
where $\alpha^a_{\bm{x}}\equiv \alpha^a({\bm{x}})$,
${\cal K}_{\bm{x} \bm{y} \bm{z}}\equiv {\cal K}({\bm{x}, \bm{y}, \bm{z}})$, and
the factor of $N_c$ has been obtained as $(T^cT^d)_{ab}\delta^{cd} = N_c \delta^{ab}$.
The first term within the brackets, proportional to ${\cal K}_{\bm{x} \bm{y} \bm{z}}$,
represents the ``real gluon'' contribution to the evolution, while the other terms make up the
``virtual'' contribution, and have been generated when commuting one of the functional derivatives
in Eq.~(\ref{RGE}) through $\eta$.

Consider first the convergence properties of the above integral over $\bm z$. As 
expected from the general discussion, 
there is no singularity at short distances :
the three terms within the integrand are separately ultraviolet finite. To study the
large distance behavior ($z \gg x,y$), it is convenient to group separately the
terms involving $\alpha^a_{\bm{z}}$, and those without it. The latter combine to
(cf. Eq.~(\ref{calK})) :
\be
\big\langle \alpha^a_{\bm{x}} \alpha^a_{\bm{y}}\big\rangle_Y
\Big\{2{\cal K}_{\bm{x} \bm{y} \bm{z}} - {\cal K}_{\bm{x} \bm{x} \bm{z}}
- {\cal K}_{\bm{y} \bm{y} \bm{z}}\Big\}\,=\,- \frac{1}{(2\pi)^2}\,
   \frac{(\bm{x}-\bm{y})^2}{(\bm{x}-\bm{z})^2 (\bm{z}-\bm{y})^2}\,
\big\langle \alpha^a_{\bm{x}} \alpha^a_{\bm{y}}\big\rangle_Y,\ee
where the $\bm{z}$--dependent prefactor decays like $1/z^4$ at large $z$, and the
ensuing integral is convergent. Note that the potentially troublesome
terms behaving like $1/z^2$ have cancelled in the 
linear combination within the braces
(i.e., in between ``real'' and ``virtual'' contributions).
But for the terms involving one or two factors of $\alpha^a_{\bm{z}}$, 
there is no such a cancellation. For instance:
\beq\label{xz}
{\cal K}_{\bm{x} \bm{y} \bm{z}} \,
\big\langle \alpha^a_{\bm{x}} \alpha^a_{\bm{z}}\big\rangle_Y \, \,
\propto  \, \,\frac{1}{z^2}\,\big\langle \alpha^a_{\bm{x}} \alpha^a_{\bm{z}}\big\rangle_Y
\qquad{\rm for}\qquad z \,\gg \, x, \,y \,,\eeq
which leaves the place for a potential infrared problem.
The only way to avoid this problem is that the 2--point function 
$\big\langle \alpha^a_{\bm{x}} \alpha^a_{\bm{z}}\big\rangle_Y$ decreases
sufficiently fast at large separations $|\bm{z}-\bm{x}|$. There is no
reason to expect such a property to hold in general. But it {\it does} hold 
for the case of interest here, i.e., for $Y=0$ and the  2--point function in Eq.~(\ref{2point0}).
 This is so because the color field of a dipole
is rapidly decreasing with the separation from the center of the dipole :
\be
{\cal G}({\bm z}|{\bm{x}}_0,{\bm{y}}_0)\,\simeq\,\frac{1}{2\pi}\,\frac{\bm{r}_0\cdot
{\bm z}}{{\bm z}^2}\qquad{\rm for}\qquad z \,\gg \, x_0, \,y_0 \,,\ee
where $\bm{r}_0=\bm{x}_0-\bm{y}_0$. Thus, when $z$ is very large, the function in
Eq.~(\ref{xz}) decays like $1/z^3$, which is rapid enough for the convergence of the integral
in Eq.~(\ref{2pointEVOL}).

To conclude, the following equation is well defined:
\be\label{2pEVOL}
\frac{{\rm d}}{{\rm d}Y}\,\big\langle \alpha^a_{\bm{x}} \alpha^a_{\bm{y}}\big\rangle_Y\bigg|_{Y=0}
&=&\frac{g^2N_c }{2\pi} \times g^2C_F\int_{\bm z}\Big\{
2{\cal K}_{\bm{x} \bm{y} \bm{z}}\big({\cal G}_{\bm{x}} - {\cal G}_{\bm{z}}\big)
({\cal G}_{\bm{y}} - {\cal G}_{\bm{z}}\big) \nn 
&{}& \qquad\qquad- \,{\cal K}_{\bm{x} \bm{x} \bm{z}}\,
{\cal G}_{\bm{y}}\big({\cal G}_{\bm{x}} - {\cal G}_{\bm{z}}\big)
- {\cal K}_{\bm{y} \bm{y} \bm{z}}\,{\cal G}_{\bm{x}}({\cal G}_{\bm{y}} - {\cal G}_{\bm{z}}\big)
\Big\},\ee
(with the shorthand notation ${\cal G}_{\bm{x}}\equiv
{\cal G}({\bm x}|{\bm{x}}_0,{\bm{y}}_0)$), so it is meaningful to perform an integration
by parts over $\bm{z}$ in its right hand side. With this aim, we shall use the second equality in
Eq.~(\ref{calK}) which shows that, when ${\cal K}_{\bm{x} \bm{y} \bm{z}}$ multiplies a
function which vanishes at $\bm{z}=\bm{x}$ and $\bm{z}=\bm{y}$ (as in
the integrand in Eq.~(\ref{2pEVOL})), it can be replaced by:
\be\label{K1}
{\cal K}_{\bm{x} \bm{y} \bm{z}}\,\longrightarrow\, \frac{1}{2}\,
\grad^2_{\bm{z}}\big[\Delta({\bm{x}}-{\bm{z}})\Delta({\bm{y}}-{\bm{z}})\big].\ee
It is tempting to use this replacement to perform the integration by parts. 
However, if one does so, one generates terms like $\grad^2_{\bm{z}}{\cal G}_{\bm{z}}^2$
which are ill defined because of singularities at $\bm{z}=\bm{x}_0$ and $\bm{z}=\bm{y}_0$.
For instance:
\be\label{M0}
\grad^i_{\bm{z}}{\cal G}_{\bm{z}}\grad^i_{\bm{z}}{\cal G}_{\bm{z}}&\equiv&
\big(\grad^i_{\bm{z}}{\cal G}({\bm z}|{\bm{x}}_0,{\bm{y}}_0)\big)^2\,=\,\big(
\grad^i_{\bm{z}}\Delta({\bm{z}}-{\bm{x}}_0)-\grad^i_{\bm{z}}\Delta({\bm{z}}-{\bm{y}}_0)\big)^2\nn
&=&  \frac{1}{(2\pi)^2}\,
\frac{(\bm{x}_0-\bm{y}_0)^2}{(\bm{x}_0-\bm{z})^2 (\bm{y}_0-\bm{z})^2}\,\ee
has uncompensated poles at $\bm{z}=\bm{x}_0$ and $\bm{z}=\bm{y}_0$.

To avoid this problem, we shall use the following trick, whose physical significance should
become clear later: We first identically rewrite Eq.~(\ref{calK}) as
\be\label{K2}
{\cal K}_{\bm{x} \bm{y} \bm{z}}&=&\frac{1}{2}\,\grad^i_{\bm{z}}\big(\Delta({\bm{x}}-{\bm{z}})
- \Delta({\bm{x}}-{\bm{x}}_0)\big)\grad^i_{\bm{z}}\big(\Delta({\bm{y}}-{\bm{z}})
- \Delta({\bm{y}}-{\bm{x}}_0)\big)\nn
&+&\frac{1}{2}\,\grad^i_{\bm{z}}\big(\Delta({\bm{x}}-{\bm{z}})
- \Delta({\bm{x}}-{\bm{y}}_0)\big)\grad^i_{\bm{z}}\big(\Delta({\bm{y}}-{\bm{z}})
- \Delta({\bm{y}}-{\bm{y}}_0)\big)\nn
&-&\frac{1}{2}\,\grad^i_{\bm{z}}\big(\Delta({\bm{x}}-{\bm{x}}_0)
- \Delta({\bm{x}}-{\bm{y}}_0)\big)\grad^i_{\bm{z}}\big(\Delta({\bm{y}}-{\bm{x}}_0)
- \Delta({\bm{y}}-{\bm{y}}_0)\big)\,.\ee
This is indeed the same as Eq.~(\ref{calK}) since the new terms included within the braces
are independent of $\bm{z}$, and thus give zero when acted on by the derivatives.
This equation too can be written as a total derivative when multiplying a
function which vanishes at $\bm{z}=\bm{x}$ and $\bm{z}=\bm{y}$ :
\be\label{Keff}\hspace*{-.9cm}
{\cal K}_{\bm{x} \bm{y} \bm{z}}\longrightarrow \frac{1}{4}
\grad^2_{\bm{z}}\Big\{{\cal G}({\bm x}|{\bm{x}}_0,{\bm{z}}){\cal G}({\bm y}|{\bm{x}}_0,{\bm{z}})
+ {\cal G}({\bm x}|{\bm{z}},{\bm{y}}_0){\cal G}({\bm y}|{\bm{z}},{\bm{y}}_0)
-{\cal G}({\bm x}|{\bm{x}}_0,{\bm{y}}_0){\cal G}({\bm y}|{\bm{x}}_0,{\bm{y}}_0)\Big\},\nn\ee
since ${\cal G}({\bm x}|{\bm{x}}_0,{\bm{z}})= \Delta({\bm{x}}-{\bm{x}}_0)-\Delta({\bm{x}}-{\bm{z}})$,
so that $\grad^2_{\bm{z}}{\cal G}({\bm x}|{\bm{x}}_0,{\bm{z}})=-\grad^2_{\bm{z}}
\Delta({\bm{x}}-{\bm{z}})=\delta^{(2)}({\bm{x}}-{\bm{z}})$, etc.
But as compared to Eq.~(\ref{K1}), the equation above has the advantage that the function within
the braces vanishes, by construction, at $\bm{z}=\bm{x}_0$ and $\bm{z}=\bm{y}_0$, so,
after integration by parts, it will compensate the singularities at these points.

Specifically, after inserting Eq.~(\ref{Keff}) into Eq.~(\ref{2pEVOL}) and performing the
integration by parts, all the terms but one --- 
namely $\big(\grad^i_{\bm{z}}{\cal G}_{\bm{z}}\big)^2$
--- cancel out, and we are left with:
\be\label{2pfinal}
\frac{{\rm d}}{{\rm d}Y}\,\big\langle \alpha^a_{\bm{x}} \alpha^a_{\bm{y}}\big\rangle_Y\bigg|_{Y=0}
\!\!&=&\!\!\frac{g^2N_c }{2\pi} \times g^2C_F\int_{\bm z} {\cal M}({\bm{x}}_0,{\bm{y}}_0,{\bm z})
\Big\{-{\cal G}({\bm x}|{\bm{x}}_0,{\bm{y}}_0){\cal G}({\bm y}|{\bm{x}}_0,{\bm{y}}_0)\nn
&{}&\qquad +\,\,{\cal G}({\bm x}|{\bm{x}}_0,{\bm{z}}){\cal G}({\bm y}|{\bm{x}}_0,{\bm{z}})
+ {\cal G}({\bm x}|{\bm{z}},{\bm{y}}_0){\cal G}({\bm y}|{\bm{z}},{\bm{y}}_0)\Big\},\ee
where the ``dipole kernel''
\be\label{Mdef}
{\cal M}({\bm{x}}_0,{\bm{y}}_0,{\bm z})\,\equiv\, \frac{1}{(2\pi)^2}\,
\frac{(\bm{x}_0-\bm{y}_0)^2}{(\bm{x}_0-\bm{z})^2 (\bm{y}_0-\bm{z})^2}\,\ee
has been generated as in Eq.~(\ref{M0}). Note that the surviving term (which gave rise
to the dipole kernel) comes from the ``real'' piece in Eq.~(\ref{2pointEVOL}) alone.
But although the ``virtual'' piece has not given any explicit contribution to the final
equation (\ref{2pfinal}), its presence in the original equation (\ref{2pointEVOL}) was nevertheless
essential to ensure infrared finiteness, and thus permit the manipulations leading to 
Eq.~(\ref{2pfinal}). The latter is manifestly infrared finite because of the rapid decay of
the dipole kernel at large distances: ${\cal M}({\bm{x}}_0,{\bm{y}}_0,{\bm z})\sim 1/z^4$
when $z\gg x_0,\,y_0$.

\begin{figure}
  \centerline{
  \epsfsize=0.99\textwidth
\epsfbox{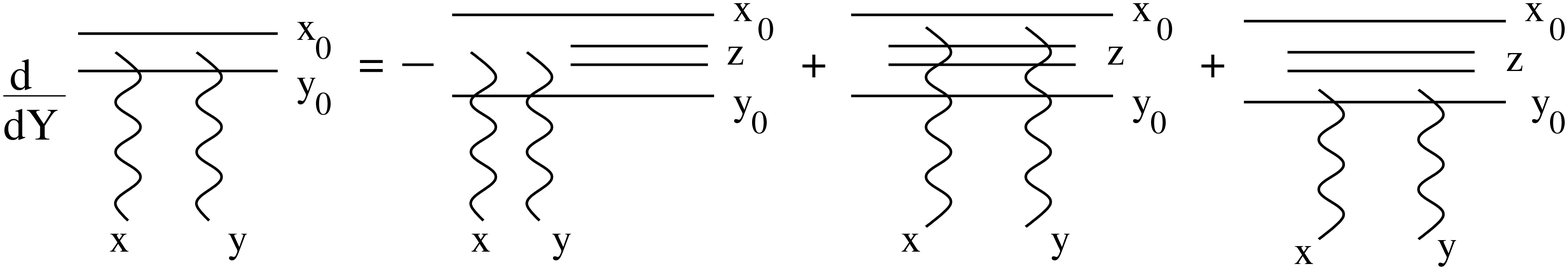}
  }
 \caption[]{Graphical illustration of Eq.~(\ref{2pfinal}).}
\label{SPLIT}
\end{figure}

Although equivalent to Eq.~(\ref{2pEVOL}), as shown by the calculations above,
Eq.~(\ref{2pfinal}) has the advantage to shed more direct light on the physical mechanism
behind the quantum evolution of the dipole: According to Eq.~(\ref{2point0}), the quantity
$g^2C_F {\cal G}({\bm x}|{\bm{x}}_0,{\bm{z}}){\cal G}({\bm y}|{\bm{x}}_0,{\bm{z}})$ is recognized
as the 2--point function $\langle \alpha^a_{\bm{x}} \alpha^a_{\bm{y}}\rangle_0$ of the color
field generated by a dipole with the quark at ${\bm{x}}_0$ and the antiquark at $\bm{z}$,
and similarly for the other terms within the braces in Eq.~(\ref{2pfinal}). Furthermore,
$(g^2N_c/2\pi){\cal M}({\bm{x}}_0,{\bm{y}}_0,{\bm z})$ is the probability per unit rapidity
for the emission of a soft gluon with transverse position ${\bm z}$
from a dipole with coordinates $({\bm{x}}_0,{\bm{y}}_0)$. Equivalently, this is also the
probability for the splitting of the original dipole $({\bm{x}}_0,{\bm{y}}_0)$ into two
new dipoles with coordinates $({\bm{x}}_0,{\bm{z}})$ and $({\bm{z}},{\bm{y}}_0)$, respectively
\cite{AM94}. Thus, Eq.~(\ref{2pfinal}) has a transparent physical interpretation, which is
illustrated in Fig.~\ref{SPLIT}: The change in the field--field correlator is due to the
splitting of the original dipole, which brings in new contributions to the field
from the produced dipoles, but subtracts a corresponding contribution of the decaying
dipole.

This evolution can be translated into a change in the weight function, which,
after one step in the quantum evolution, takes the following form (compare to 
Eq.~(\ref{DWF})) :
\be\label{WFdY}\hspace*{-.5cm}
W_{{\rm d}Y}[\alpha]\!\!&=&\!\!P_1({\rm d}Y)
\int \prod_{a=1}^{N_g}dQ^a\,\,{\rm e}^{-\frac{Q^aQ^a}{2\lambda}}
\,\delta\big[\alpha^a({\bm{x}}) - Q^a {\cal G}({\bm x}|{\bm{x}}_0,{\bm{y}}_0)\big]\\
&+&\!\!\int_{\bm z} P_2({\bm{z}}|{\rm d}Y)
\int \prod_{i=1,2}\prod_{a=1}^{N_g}dQ^a_i\,\,{\rm e}^{-\frac{Q^a_iQ^a_i}{2\lambda}}
\,\delta\big[\alpha^a({\bm{x}}) - Q^a_1 {\cal G}({\bm x}|{\bm{x}}_0,{\bm{z}})
- Q^a_2{\cal G}({\bm x}|{\bm{z}},{\bm{y}}_0)\big],\nonumber\ee
with the probabilities:
\be\label{PdY}
P_1({\rm d}Y)&=&1 - {\rm d}Y\,\frac{g^2N_c }{2\pi}
\int_{\bm z} {\cal M}({\bm{x}}_0,{\bm{y}}_0,{\bm z}),\nn
P_2({\bm{z}}|{\rm d}Y)&=&{\rm d}Y\,\frac{g^2N_c }{2\pi}\,{\cal M}({\bm{x}}_0,{\bm{y}}_0,{\bm z}).\ee
Note that the integral over ${\bm z}$ in the formula for $P_1$ has logarithmic singularities
at $\bm{z}=\bm{x}_0$ and $\bm{z}=\bm{y}_0$, and thus must be computed with an ultraviolet cutoff
(a minimal distance $\rho$ : $|\bm{z}-\bm{x}_0| >\rho$ and $|\bm{z}-\bm{y}_0| >\rho$).
Such singularities are to be expected in the wave function --- since there is an infinite
probability to emit an arbitrarily small dipole ---, but they cancel out in physical quantities,
as manifest, e.g., on the r.h.s. of Eq.~(\ref{2pfinal}), which is free of any (ultraviolet or
infrared) problem, as explained before. Also, they cancel in the condition of probability
conservation, which at this stage reads:
\be
P_1({\rm d}Y)\,+\,\int_{\bm z} P_2({\bm{z}}|{\rm d}Y)\,=\,1.\ee

Eq.~(\ref{WFdY}) has indeed the structure anticipated in Eq.~(\ref{WFY}). In the next subsection,
we shall prove this structure for arbitrary $Y$ (within the validity range of the weak 
field approximation).

\subsection{The BFKL evolution of the onium weight function}

To discuss the general case, it is more convenient to use an alternative form of
the weight function (\ref{WFY}), in which the average over color is explicitly performed.
As we shall see, this new form is a generalization of Eq.~(\ref{DWFexp}), but unlike the
latter it cannot be obtained via the straightforward perturbative expansion of the 
$\delta$--functional in Eq.~(\ref{WFY}). The reason is that, for our subsequent study
of unitarization, we are interested in relatively high energies, where the average number
of gluons in the onium wavefunction $N(Y)$ is so large, $N(Y)\sim 1/\alpha_s$, that it
interferes with the perturbative expansion.
Note that this is not in contradiction with the weak field assumption that we have
used so far: For the strong field effects, like saturation, to be important,
the dipole number should be even larger\footnote{This condition can be also stated
in terms of the strength of the typical fluctuations of the field $\alpha$ :
the strong field regime corresponds to $g\alpha \sim 1$ (cf. Sect. 3.2), or, for
the 2--point function: $\alpha_s \langle\alpha\alpha\rangle \sim 1$. Since
$\langle\alpha\alpha\rangle_Y \sim \alpha_s N(Y)$, as it should be clear from Eq.~(\ref{WFY}),
this gives again the condition $\alpha_s^2 N(Y)\sim 1$. By contrast, when
$\alpha_s N(Y)\sim 1$, the field amplitude is of order one, and $g\alpha$ is truly
perturbative.}, namely, $N(Y)\sim 1/\alpha_s^2$. (Indeed,
the probability that a given dipole within the onium interact with any one of
the other dipoles is of order $\alpha_s^2 N(Y)$.) Still, already when $\alpha_s N(Y)\sim 1$,
the perturbative expansion needs to be reorganized, to permit the resummation of the terms of
order $(\alpha_sN(Y))^n$ with $n\ge 1$.

To illustrate the difficulty, consider just the second--order term in the expansion of the 
$\delta$--functional in Eq.~(\ref{WFY}). After performing the average over color, this
term is formally of order $\alpha_s$ (with the shorthand notation ${\cal G}_i({\bm{x}})\equiv
{\cal G}({\bm x}|{\bm{x}}_i,{\bm{y}}_i)$) :
\be\label{2nd}
\frac{g^2}{4N_c}\sum_{i=1}^N\int_{{\bm u},{\bm v}}\,{\cal G}_i({\bm{u}}){\cal G}_i({\bm{v}})
\,\frac{\delta^2}{\delta \alpha^a({\bm{u}})
\delta \alpha^a({\bm{v}})}\,\,\delta[\alpha]\,.\ee
In reality, however, this contribution is of order $\alpha_s N$, because of the $N$ terms in the sum,
and thus is of order one for a typical configuration with $N\sim N(Y)  \sim 1/\alpha_s$.
Similarly, there are higher terms in this expansion which are of order $(\alpha_s N)^n$,
and thus contribute to leading order too. In order to isolate these terms, and perform the color
average for them, we shall use the integral representation of the 
$\delta$--functional :
\be
\delta\bigg[\alpha^a - \sum_{i=1}^N Q^a_i {\cal G}_i\bigg]\,=\,\int \! {\rm D}[\xi]\,\,
{\rm e}^{\,{\rm i}\int_{\bm x} \xi^a({\bm x})\alpha^a({\bm x})}\,
{\rm e}^{\,-{\rm i}\sum_{i=1}^N Q^a_i \int_{\bm x} \xi^a({\bm x}){\cal G}_i({\bm x})}\,.\ee
The Gaussian integrations over the color charges $Q^a_i$ are now easily performed, to give:
\be\hspace*{-.6cm}
\bigg\langle\delta\bigg[\alpha^a - \sum_{i=1}^N Q^a_i {\cal G}_i\bigg]\bigg\rangle_Q
=\int\! {\rm D}[\xi]\,\,{\rm e}^{\,{\rm i}\int_{\bm x} \xi^a({\bm x})\alpha^a({\bm x})}\,
\prod_{i=1}^N {\rm exp}\bigg\{-\frac{\lambda}{2}\bigg(\int_{\bm x} \xi^a({\bm x}){\cal G}_i({\bm x})
\bigg)^2\bigg\}.\ee
For each dipole $i$, the exponent $\lambda(\int_{\bm x} \xi{\cal G}_i)^2$ is truly a small
quantity, of order $\lambda\equiv g^2/2N_c$, so it is legitimate to preserve just the second
order term in each exponential. After this expansion, the functional 
integral over the auxiliary variables can be also performed : 
\be\hspace*{-.6cm}
\bigg\langle\delta\bigg[\alpha^a - \sum_{i=1}^N Q^a_i {\cal G}_i\bigg]\bigg\rangle_Q\,
\approx\,\,
\prod_{i=1}^N \,\bigg\{1+ \frac{\lambda}{2}\bigg(\int_{\bm x} {\cal G}_i({\bm x})
\frac{\delta}{\delta \alpha^a({\bm{x}})}\bigg)^2\bigg\}\,\delta[\alpha]\,.\ee

Thus, the expression for the weight function that we shall use in the remaining part of this
section, and which is equivalent to Eq.~(\ref{WFY}) to the accuracy of interest, reads:
\be\label{WFYexp}
W_Y[\alpha]=\sum_{N=1}^\infty\int\! d\Gamma_N\,\,P_N(Y)\,
\prod_{i=1}^N \bigg\{1+ \frac{\lambda}{2}
\int_{{\bm u},{\bm v}}\,{\cal G}_i({\bm{u}}){\cal G}_i({\bm{v}})
\,\frac{\delta^2}{\delta \alpha^a({\bm{u}})
\delta \alpha^a({\bm{v}})}\bigg\}\,\,\delta[\alpha]\,.\ee
We would like to show that this expression is a solution to the RGE (\ref{RGE})
with kernel (\ref{etalin}), and deduce in the process the evolution equations for the
probabilities $P_N(Y)$. To that aim, we shall work with the evolution equation in the form
(\ref{Oevol}), where ${\cal O}[\alpha]$ is an arbitrary operator; this equation is convenient
since it involves a  functional integral over $\alpha$, which facilitates the work with the
functional derivatives in Eq.~(\ref{WFYexp}).

After inserting the expression (\ref{etalin}) for $\eta$, the r.h.s. of Eq.~(\ref{Oevol})
can be decomposed into two pieces: a ``real'' piece, in which both functional derivatives
act on ${\cal O}[\alpha]$, and a ``virtual'' piece, which is generated when one of these
derivatives act on $\eta$. As for the example of the 2--point function discussed in Sect. 3.2,
the  ``virtual''piece is important to cancel infrared singularities within the ``real'' piece,
but it gives no explicit contribution to the final equation that we shall establish.
Therefore, in what follows we shall restrict ourselves to the ``real'' contribution, which reads:
\be\label{Oreal}
\frac{\del \langle {\cal O}\rangle_Y}{\del Y}&=&\frac{g^2}{2\pi}\big(T^c T^d\big)_{ab}
\sum_{N=1}^\infty\int d\Gamma_N\,\,P_N(Y)\nn
&{}&\times\,\,\int\! {\rm D}[\alpha]\,\delta[\alpha]\,
\prod_{i=1}^N \bigg\{1+ \frac{\lambda}{2}
\int_{{\bm u},{\bm v}}\,{\cal G}_i({\bm{u}}){\cal G}_i({\bm{v}})
\,\frac{\delta^2}{\delta \alpha^e_{\bm{u}}
\delta \alpha^e_{\bm{v}}}\bigg\}\,\nn
&{}&\times\,\,\int_{{\bm{x}, \bm{y}, {\bm z}}}
 {\cal K}({\bm{x}, \bm{y}, \bm{z}})\,\big[
\alpha^c_{\bm{x}} - \alpha^c_{\bm{z}}\big]\big[
\alpha^d_{\bm{y}} - \alpha^d_{\bm{z}}\big]\,{\delta^2 {\cal O}[\alpha]
\over {\delta \alpha^a_{\bm{x}} \delta \alpha^b_{\bm{y}}}}\,\,,\ee
after performing some integrations by parts in the functional integral.

The next step is to compute the action of the functional derivatives coming from the
weight function. We shall do this in the large--$N_c$ limit, to simplify the color algebra.
Because of the presence of $\delta[\alpha]$ in the functional integral, it is clear that
two of the functional derivatives must act on the explicit quadratic form in $\alpha$.
There are two possibilities: (i) both these derivatives refer to the same dipole, and (ii)
they refer to different dipoles. However, the second possibility is
suppressed at large $N_c$ (since, unlike the first type of contribution, it does
not provide a factor of $N_c$), and will be neglected in what follows. 
We thus obtain:
\be\label{O1}
\frac{\del \langle {\cal O}\rangle_Y}{\del Y}&=&\lambda\times \frac{g^2 N_c}{2\pi}
\sum_{N=1}^\infty\int d\Gamma_N\,\,P_N(Y)\nn
&{}&\times\,\,\sum_{i=1}^N \int_{{\bm{x}, \bm{y}, {\bm z}}}
 {\cal K}({\bm{x}, \bm{y}, \bm{z}})\,\big[{\cal G}_i({\bm{x}})-
{\cal G}_i({\bm{z}})\big]\big[{\cal G}_i({\bm{y}})-{\cal G}_i({\bm{z}})\big]\,\nn
&{}&\times\,\,\prod_{j\ne i} \bigg\{1+ \frac{\lambda}{2}
\int_{{\bm u},{\bm v}}\,{\cal G}_j({\bm{u}}){\cal G}_j({\bm{v}})
\,\frac{\delta^2}{\delta \alpha^e_{\bm{u}}
\delta \alpha^e_{\bm{v}}}\bigg\}
\,{\delta^2 {\cal O}[\alpha]
\over {\delta \alpha^a_{\bm{x}} \delta \alpha^a_{\bm{y}}}}\bigg|_{\alpha=0}\,\,.\ee

Next, we shall perform an integration by parts in the integral over $\bm{z}$ in the
second line of Eq.~(\ref{O1}). Note the similitude between this integral and that
involving the ``real'' piece in Eq.~(\ref{2pEVOL}). Like in that case, the function
${\cal G}_i({\bm{z}})\equiv \Delta(\bm{z} - \bm{x}_i) - \Delta(\bm{z} - \bm{y}_i) $
has logarithmic singularities at $\bm{z} =\bm{x}_i$ and $\bm{z} =\bm{y}_i$, which
would be amplified by a too ``brutal'' integration by parts, based on Eq.~(\ref{K1}).
Once again, this difficulty can be avoided by using the analog of Eq.~(\ref{Keff})
(here, separately for each dipole) :
\be\label{Keffi}\hspace*{-.9cm}
{\cal K}_{\bm{x} \bm{y} \bm{z}}\longrightarrow \frac{1}{4}
\grad^2_{\bm{z}}\Big\{{\cal G}({\bm x}|{\bm{x}}_i,{\bm{z}}){\cal G}({\bm y}|{\bm{x}}_i,{\bm{z}})
+ {\cal G}({\bm x}|{\bm{z}},{\bm{y}}_i){\cal G}({\bm y}|{\bm{z}},{\bm{y}}_i)
-{\cal G}({\bm x}|{\bm{x}}_i,{\bm{y}}_i){\cal G}({\bm y}|{\bm{x}}_i,{\bm{y}}_i)\Big\},\nn\ee
where the expression within the braces vanishes at $\bm{z} =\bm{x}_i$ and $\bm{z} =\bm{y}_i$.
Still as in Sect. 3.2, the integration by parts generates the ``dipole kernel'' (for
the $i$th dipole) :
\be\label{Mi}
{\cal M}({\bm{x}}_i,{\bm{y}}_i,{\bm z})\,\equiv\,\big (
\bfgrad_{\bm z} \,{\cal G}_i({\bm{z}})\big)^2
\,=\, \frac{1}{(2\pi)^2}\,
\frac{(\bm{x}_i-\bm{y}_i)^2}{(\bm{x}_i-\bm{z})^2 (\bm{y}_i-\bm{z})^2}\,.\ee
The three terms within the braces in  Eq.~(\ref{Keffi}) are then naturally combined
with the functional derivatives acting on ${\cal O}[\alpha]$, to finally yield:
\be\label{Ofinal}
\frac{\del \langle {\cal O}\rangle_Y}{\del Y}&=& \frac{g^2 N_c}{2\pi}
\sum_{N=1}^\infty\int d\Gamma_N\,\,P_N(Y)\,\sum_{i=1}^N 
\,\prod_{j\ne i} \bigg\{1+ \frac{\lambda}{2}\bigg(
\int_{\bm u}\,{\cal G}_j({\bm{u}}) \frac{\delta}{\delta \alpha^a_{\bm{u}}} \bigg)^2\bigg\}\nn
&{}&\quad\times\,\,\int_{\bm z} {\cal M}({\bm{x}}_i,{\bm{y}}_i,{\bm z})
\bigg\{- \frac{\lambda}{2}\bigg(\int_{\bm x}\,{\cal G}({\bm x}|{\bm{x}}_i,{\bm{y}}_i)
\frac{\delta}{\delta \alpha^a_{\bm{x}}} \bigg)^2 \\
&{}&\qquad\qquad + \,\,
\frac{\lambda}{2}\bigg(\int_{\bm x}\,{\cal G}({\bm x}|{\bm{x}}_i,{\bm{z}})
\frac{\delta}{\delta \alpha^a_{\bm{x}}} \bigg)^2
+ \frac{\lambda}{2}\bigg(\int_{\bm x}\,{\cal G}({\bm x}|{\bm{z}},{\bm{y}}_i)
\frac{\delta}{\delta \alpha^a_{\bm{x}}} \bigg)^2
 \bigg\}\,{\cal O}[\alpha]\bigg|_{\alpha=0}\,.\nonumber \ee

Since the operator ${\cal O}[\alpha]$ is arbitrary, the equation above is equivalent to an
equation for $\del W_Y/\del Y$, which is most suggestively written for
$W_{Y+{\rm d}Y}=W_Y + (\del W_Y/\del Y){\rm d}Y$ :
\be\label{WYdY}
\hspace*{-1.cm}W_{Y+{\rm d}Y}[\alpha] = \sum_{N=1}^\infty\int d\Gamma_N\,\,P_N(Y)\,
\qquad\qquad\qquad\qquad\qquad\qquad\qquad\qquad\qquad\qquad\qquad\qquad\qquad\qquad\\
\times\,\Bigg\{\bigg[1 - {\rm d}Y\,\frac{g^2N_c }{2\pi} \sum_{i=1}^N 
\int_{\bm z} {\cal M}({\bm{x}}_i,{\bm{y}}_i,{\bm z})\bigg]
\prod_{j=1}^N \bigg[1+ \frac{\lambda}{2}\bigg(
\int_{\bm x}\,{\cal G}_j({\bm{x}}) \frac{\delta}{\delta \alpha^a_{\bm{x}}} \bigg)^2\bigg]
\qquad\qquad\quad\nn
\quad +\,\,{\rm d}Y\,\frac{g^2N_c }{2\pi}\, \sum_{i=1}^N \,\prod_{j\ne i} 
\bigg[1+ \frac{\lambda}{2}\bigg(
\int_{\bm x}\,{\cal G}_j({\bm{x}}) \frac{\delta}{\delta \alpha^a_{\bm{x}}} \bigg)^2\bigg]
\qquad\qquad\qquad\qquad\qquad\qquad\nn
\times\,\int_{\bm z} {\cal M}({\bm{x}}_i,{\bm{y}}_i,{\bm z})
\bigg[1+\frac{\lambda}{2}\bigg(\int_{\bm x}\,{\cal G}({\bm x}|{\bm{x}}_i,{\bm{z}})
\frac{\delta}{\delta \alpha^a_{\bm{x}}} \bigg)^2\bigg]
\bigg[1+\frac{\lambda}{2}\bigg(\int_{\bm x}\,{\cal G}({\bm x}|{\bm{z}},{\bm{y}}_i)
\frac{\delta}{\delta \alpha^a_{\bm{x}}} \bigg)^2\bigg]
\Bigg\}\,\delta[\alpha]\,.\nonumber\ee
After comparing this to Eq.~(\ref{WFYexp}), the effects of the evolution become
transparent: When increasing the rapidity in one step, a given $N$--dipole configuration
can either survive as it is, but with a smaller probability, or evolve by radiating one
soft gluon, which in the large--$N_c$ limit is equivalent to the splitting of one of the
$N$ original dipoles into a pair of new dipoles. To demonstrate that this evolution is
indeed consistent with Eq.~(\ref{WFYexp}), we still have to show that Eq.~(\ref{WYdY})
is of the form:
\be\label{WYdP}
W_{Y+{\rm d}Y}[\alpha]=\sum_{N=1}^\infty\int\! d\Gamma_N\,\,P_N(Y+{\rm d}Y)\,
\prod_{i=1}^N \bigg[1+ \frac{\lambda}{2}\bigg(
\int_{\bm x}\,{\cal G}_i({\bm{x}}) \frac{\delta}{\delta \alpha^a_{\bm{x}}} \bigg)^2\bigg]
\,\delta[\alpha],\ee
and thus deduce the evolution equations for the probabilities. 

For this purpose, it is convenient to adopt an economical labelling of the dipoles,
which takes into account the specificity of the quantum evolution: A newly produced
pair of dipoles involves just one more transverse coordinate in addition to the two
coordinates of the parent dipole (see e.g. Fig.~\ref{SPLIT}). 
Thus, if one starts with the dipole $(\bm{x}_0,{\bm{y}}_0)$
at $Y=0$ and follows its evolution with increasing $Y$, then one needs one more coordinate,
say $\bm{x}_1$, to describe the ensuing two dipole system, two more coordinates, 
say $\bm{x}_1$ and $\bm{x}_2$, for a three dipole system, and so on. 

Therefore, a 
$N$--dipole configuration generated by the evolution up to rapidity $Y$ can be labelled by
$N-1$ transverse coordinates $\bm{x}_1$, $\bm{x}_2,\dots,\bm{x}_{N-1}$ (physically, these
are the positions of the emitted gluons), and thus can be characterized by a
probability density $P_N(\bm{x}_1,\bm{x}_2,\dots,\bm{x}_{N-1}|Y)$. (As before, the
dependence upon the original coordinates $(\bm{x}_0,{\bm{y}}_0)$ is kept implicit.)
With this labelling, the coordinates of the $N$ dipoles are:
$(\bm{x}_0,{\bm{x}}_1),\,\,({\bm{x}}_1,\bm{x}_2),\,\,\dots\,\,(\bm{x}_{N-1},{\bm{y}}_0),$
and the measure for the phase--space integration is simply:
\be\label{GAMMAN}
d\Gamma_N\,=\,{\rm d}^2\bm{x}_1{\rm d}^2\bm{x}_2\dots{\rm d}^2\bm{x}_{N-1}\,.\ee
In particular, the $i$th dipole in the configuration is that with coordinates
$({\bm{x}}_{i-1},\bm{x}_i)$. Thus, we have to replace retrospectively $(\bm{x}_i,
\bm{y}_i)\longrightarrow ({\bm{x}}_{i-1},\bm{x}_i)$ in all the previous formulae.

We are finally in a position to check that Eq.~(\ref{WYdY}) can be indeed brought into the
form (\ref{WYdP}), and deduce that, for this to be possible, the probabilities $P_N$ must obey
the following recurrence formula:
\be\label{recurP}
\hspace*{-1.cm}P_N(\bm{x}_1,\dots,\bm{x}_{N-1}|Y+{\rm d}Y)=
\bigg[1 - {\rm d}Y\,\frac{g^2N_c }{2\pi} \sum_{i=1}^N 
\int_{\bm z} {\cal M}({\bm{x}}_{i-1},{\bm{x}}_i,{\bm z})\bigg]
P_N(\bm{x}_1,\dots,\bm{x}_{N-1}|Y)\nn
+\,\, {\rm d}Y\,\frac{g^2N_c }{2\pi} \sum_{i=1}^{N-1}
{\cal M}({\bm{x}}_{i-1},{\bm{x}}_{i+1},{\bm{x}}_i)
P_{N-1}(\bm{x}_1,\dots,\hat{\bm{x}}_i,\dots,\bm{x}_{N-1}|Y)\,.\qquad\ee
In the second line, a hat on the argument ${\bm{x}}_i$ in $P_{N-1}$ means that
the coordinate ${\bm{x}}_i$ is actually missing from the respective configuration
of $N-1$ dipoles (that is, the $i$th dipole in that configuration has coordinates
$({\bm{x}}_{i-1},\bm{x}_{i+1})$).

The r.h.s. of Eq.~(\ref{recurP}) is recognized as the sum of two terms: a loss term
and a gain term. While the loss term, which describes the emission of one gluon
from the original $N$--dipole configuration, can be easily read off Eq.~(\ref{WYdY}), 
the gain term, on the other hand, which describes the formation of the $N$--dipole 
configuration via the splitting of one dipole in an original configuration with 
only $N-1$ dipoles, is more subtle, and can be recognized only after a judicious
change of variables in the last term in Eq.~(\ref{WYdY}). This is explained in the
Appendix.

Clearly, the recurrence formula (\ref{recurP}) can be also written as a set of coupled evolution
equation for the probabilities $P_N$:
\be\label{evolP}
\frac{\del P_N(Y)}{\del Y}&=& -\,\frac{\alpha_s N_c}{2\pi^2} \bigg[\sum_{i=1}^N 
\int_{\bm z} \frac{(\bm{x}_{i-1}-\bm{x}_i)^2}{(\bm{x}_{i-1}-\bm{z})^2 (\bm{x}_i-\bm{z})^2}\bigg]
P_N(Y)\nn &{}& +\,\,\frac{\alpha_s N_c}{2\pi^2}\sum_{i=1}^{N-1} 
\frac{(\bm{x}_{i-1}-\bm{x}_{i+1})^2}{(\bm{x}_{i-1}-\bm{{x}_i})^2 (\bm{x}_{i+1}-\bm{{x}_i})^2}
\,\,P_{N-1}(\hat{\bm{x}}_i|Y),\ee
where we have omitted all the obvious arguments.
These equations must be solved with initial conditions which follow from Eq.~(\ref{DWFexp}),
namely $P_N(Y=0)=\delta_{N1}$.

It is straightforward to check that the probability is correctly conserved by
the evolution (\ref{evolP}) (see the Appendix) :
\be\label{PROBY}
\sum_{N=1}^\infty\int  d\Gamma_N\,\,P_N(Y)\,=\,1.\ee
In fact, because of the simple structure of the equations (\ref{evolP}), in which $P_N$
is coupled only to $P_{N-1}$, the  probability conservation is even more stringent,
in the sense that it holds already for configurations with neighbouring numbers of dipoles:
\be\label{PROBN}
\int  d\Gamma_N\,\,\frac{\del P_N^{(l)}}{\del Y}\,+\,\int  d\Gamma_{N+1}
\,\,\frac{\del P_{N+1}^{(g)}}{\del Y}\,=\,0.\ee
In this equation, which will be proven in the Appendix, $\del P_N^{(l)}/\del Y$
is the loss term in Eq.~(\ref{evolP}), while $\del P_{N+1}^{(g)}/\del Y$ is the gain
term in the corresponding equation for $P_{N+1}$.

The coupled equations (\ref{evolP}) can be solved iteratively for successively higher
probabilities: For $N=1$, one has a closed equation for $P_1(Y)$; 
once this is solved, its solution $P_1(Y)$ is 
inserted in the equation with $N=2$ to give a closed equation for $P_2(Y)$, etc.
But in order for these equations to be well defined, they must be supplemented
with an ultraviolet cutoff: Indeed, written as it stands, 
the integral over $\bm{z}$ in the loss term is afflicted with logarithmic singularities 
due to the poles of the integrand at $\bm{z}= \bm{x}_i$ and $\bm{z}= \bm{x}_{i-1}$.
These singularities reflect the fact that we cannot forbid the radiation of 
dipoles of arbitrarily small sizes. Rather, we shall require a minimal size 
$\rho$ for the radiated dipoles, which plays the role of an ultraviolet cutoff,
and upon which the probabilities $P_N$ depend logarithmically.
(For instance, $P_1(Y;\rho)$ represents the probability that, 
after the evolution through the rapidity interval $Y$, the onium wavefunction 
contains no more dipoles of size larger than $\rho$ other then the parent dipole.)
But this cutoff dependence cancels out in the calculation of any physical quantity, 
which involves a sum over all probabilities. 
This is clear, for instance, from Eq.~(\ref{Ofinal}), which governs the
evolution of an arbitrary observable: in the r.h.s. of that equation, the integral over
the coordinate $\bm{z}$ of the newly emitted gluon is free of short--range (and
also long--range) singularities, since, e.g., ${\cal G}({\bm x}|{\bm{x}}_i,{\bm{z}})\to 0$
when $\bm{z}\to \bm{x}_i$. A more specific example will be given in the Appendix,
where we shall construct the evolution equation for the dipole number density
(which turns out to be the BFKL equation), 
and demonstrate in the process the cancellation of ultraviolet singularities between
the loss and gain terms.

To summarize, the weight function (\ref{WFYexp}) (or (\ref{WFY})), with the probabilities
$P_N(Y)$ evolving according to the linear system of equations (\ref{evolP}), is the solution
to the RGE in the weak field, or BFKL, regime and in the large--$N_c$ approximation.
This includes the regime in which the total number of dipoles (or gluons) is as large as 
$N(Y)\sim 1/\alpha_s$, which corresponds to a {\it gluon distribution} ${\rm d}N/{\rm d}Y$
of order one. But this fails to apply at energies so high that $N(Y)\sim 1/\alpha_s^2$
(i.e., the gluon distribution is of order $1/\alpha_s$), where
the saturation effects are important.

These are precisely the assumptions used in Refs. \cite{AM94,AM95} to construct
the onium wavefunction in the color dipole picture. Thus, it makes sense to compare
the two pictures, and after doing so, it turns out that they are equivalent indeed.
This is most easily seen by comparing the previous results in this section with
the version of CDP used by Salam in his Monte-Carlo simulations of the
onium wavefunction: In Ref.\cite{Salam95}, Salam has used a recurrence formula 
equivalent to Eq.~(\ref{recurP}) to numerically construct the onium.

\section{Onium--onium scattering at high energies}
\setcounter{equation}{0}

We now have all the ingredients necessary to study the 
onium--onium scattering within the color glass formalism, up to energies
which are high enough for the unitarization effects to play a role.
This requires putting together the onium weight function $W_Y[\alpha]$ that
we have constructed in the previous section, and the factorized formula for 
the elastic scattering between two color glasses that we have proposed in Sect. 2.

\subsection{Symmetric scattering of two color glasses}

More precisely, in Sect. 2 we have shown that the symmetric formula (\ref{CGCDD})
can be used to describe the low--energy scattering between two elementary dipoles.
Here, we shall argue that this formula can be extended to high energies
provided both color glasses remain non--saturated
(that is, they remain in the weak field regime, as characterized in Sect. 3). 
Specifically, the generalization of Eq.~(\ref{CGCDD}) to high energies reads:
\be\label{SYCGC}
S_Y
\,=\,\int {\rm D}[\alpha_R]\, \,W_{Y-y}[\alpha_R]\int {\rm D}[\alpha_L]\, \,W_y[\alpha_L]
\,\,{\rm e}^{\,{\rm i}\int d^2{\bm z}\,\grad^i \alpha^a_L({\bm{z}})
\grad^i\alpha^a_R({\bm{z}})}\,,
\ee
where the two weight functions are computed in the BFKL approximation (cf. Sect. 3),
and the rapidities $Y-y$ and $y$ should be such that the weak field condition is
satisfied for the two incoming systems.
This implies a frame--dependent upper limit on the total rapidity $Y$
up to which Eq.~(\ref{SYCGC}) can be used.
Clearly, the optimal choice is the center--of--mass (CM) frame,
$y=Y-y=Y/2$, since this allows the highest value for $Y$ before 
saturation effects start to be important in any of the two color glasses.
This  highest value is determined by the condition $\alpha_s^2 N(Y/2) \sim 1$,
which, together with the BFKL estimate for the gluon number :
$N(Y) \sim {\rm e}^{\omega_0 Y}$ (with $\omega_0 = (4\ln 2)\alpha_s N_c/\pi$), provides
the following upper limit for the validity of Eq.~(\ref{SYCGC}) in the CM frame:
\be\label{Yc}
Y_c\,\simeq\,\frac{2}{\omega_0}\,\ln\,\frac{1}{\alpha_s^2}\,.\ee

That Eq.~(\ref{SYCGC}) describes correctly the single (BFKL) pomeron exchange approximation,
should be rather obvious: After expanding the exponential there to second order,
one generates the 2--point functions of the color fields in each color glass, and these
are well known to obey the BFKL equation \cite{JKLW97,PI}.
(For the particular case of onium--onium scattering, this will be verified in the Appendix.)
What is, however, more interesting
is that Eq.~(\ref{SYCGC}) can be trusted also {\it beyond}
the single--scattering approximation. That is, the multiple scattering series, as encoded
in the higher order terms in the expansion of the exponential, is consistently described
by this equation up to energies well above the onset of unitarization.
We shall not prove this in general, but simply show that, for onium--onium scattering, 
Eq.~(\ref{SYCGC}) reproduces the correct result, including multiple pomeron exchanges,
as originally obtained in the color dipole picture\cite{AM95}.

But one can nevertheless understand why, in this formulation, there is no inconsistency
between having a linear approximation for the wavefunctions and keeping non--linear
terms in the collision: The point is that, even if the color fields are indeed
weak in the individual wavefunctions, in such a way that $\alpha_s^2 N(Y/2) \ll 1$,
the scattering between the two systems can still be strong, since enhanced by
the number of gluons (or dipoles) in {\it both} systems, and thus of order 
$(\alpha_s N(Y/2))^2$. This becomes of order one already for $Y\sim Y_c/2$, showing
that there exists an interesting range of intermediate rapidities, namely,
\be\label{range}   Y_c/2\,\,\simle Y\,\,\ll\,\,Y_c\,,\ee
within which saturation effects are still negligible, but the unitarization effects
set in. Of course, the existence of such an intermediate regime is specific to
the CM frame. If we were to work in an asymmetric frame, like the rest frame of one
of the two hadrons, then unitarization effects and saturation effects in the wavefunction
of the energetic hadron would start to be important at the same energy, namely,
for $Y\sim Y_c/2$.

One may wonder, what happens if one, or both, of the color glasses
are saturated. What would be the generalization of Eq.~(\ref{SYCGC}) to that case ?
(This would be interesting, e.g., to study collisions with $Y > Y_c$ in the CM frame.)
As we know, in the CGC formalism there is no difficulty of principle of dealing
with a saturated wavefunction. This is determined by the functional RGE, which
is explicitly known in the non--linear regime (although rather tedious to solve
there). The true difficulty, which prevents us from extending the symmetric formula
(\ref{SYCGC}) to the non--linear regime at saturation, is that we do not
know how to couple a generic distribution of classical color charges to a strong
classical color field. That is, we do not know how to generalize the eikonal coupling
in Eq.~(\ref{SYCGC}) to the strong field regime in the non--Abelian case.

In fact, it is not clear whether such a generalization exists, even in principle. 
Recall from Sect. 2 that we do know how to couple a simple projectile,
like a color dipole, to a color glass condensate. This involves Wilson lines which describe
the multiple scattering of each elementary constituent in the projectile off the
 color field in the target. But this requires a detailed information
about the color matrix structure of each constituent,
which may be difficult, if not
impossible, to encode in an average description of the color glass type.
As we shall see in the following calculations, this difficulty does not appear
for onium--onium scattering in the rapidity range (\ref{range}) since an elementary 
dipole from one onium undergoes only single scattering off the color field in the other onium.

One may finally observe that using  a symmetric formula like Eq.~(\ref{SYCGC}) at
arbitrarily high energies it not really necessary. Once one accepts to work with
saturated wavefunctions, one can very well use an asymmetric frame, in which
the factorization of the $S$--matrix is better under control (recall, e.g.,
Eqs.~(\ref{Sdef})--(\ref{CGCaverage})). The advantage of the formula (\ref{SYCGC})
is precisely to extent the use of {\it non--saturated} wavefunctions up to 
energies well above the threshold for the onset of unitarization.

\subsection{Application to onium--onium scattering}

Let us now use the onium weight function in Eq.~(\ref{WFY}) to compute
the $S$--matrix element for onium--onium scattering in Eq.~(\ref{SYCGC}).
Eq.~(\ref{WFY}) is more convenient for this purpose (as compared to the other
form of the weight function, Eq.~(\ref{WFYexp})) since the functional integrals
over $\alpha_R$ and $\alpha_L$ can be immediately performed, with the result:
\be\label{SY1}\hspace*{-.7cm}
S_Y=\sum_{N=1}^\infty\int \! d\Gamma_N\,P_N(Y/2)\,
\sum_{N'=1}^\infty\int \! d\Gamma_{N'}\,P_{N'}(Y/2)\,
\bigg\langle{\rm exp}\bigg\{\,{\rm i} \sum_{i=1}^N\sum_{j=1}^{N'}Q_i^a \bar Q_j^a
{\cal D}(i|j)\bigg\}\bigg\rangle_{Q,\bar Q}.\,\,\ee
In this equation, $Q_i^a$ are color charges for the $N$ right--moving dipoles,
$\bar Q_j^a$ are the corresponding quantities for the $N'$ left--moving dipoles, and
(cf. Eq.~(\ref{calD})) :
\be
{\cal D}(i|j)\,\equiv\,{\cal D}({\bm{x}}_{i-1},{\bm{x}}_i|{\bm y}_{j-1},{\bm{y}}_j)
\ee
is a shorthand notation for the interaction potential between the right--moving dipole
with coordinates
$({\bm{x}}_{i-1},{\bm{x}}_i)$ and the left--moving one at $({\bm y}_{j-1},{\bm{y}}_j)$.
The brackets refer to the average over color, to be performed separately for each of the
$N\times N'$ dipoles (cf. Eq.~(\ref{WFY})). It is straightforward to compute this
average explicitely for one set of color variables, say, those associated with the 
right--movers. This gives:
\be\label{SY2}
S_{N\times N'}\equiv 
\bigg\langle{\rm e}^{\,{\rm i} \sum_{i=1}^N Q_i^a \sum_{j=1}^{N'} \bar Q_j^a
{\cal D}(i|j)}\bigg\rangle_{Q,\bar Q}\,=\,
\bigg\langle 
{\rm exp}\bigg\{-\frac{\lambda}{2} \sum_{i=1}^N \bigg[\sum_{j=1}^{N'}\bar Q_j^a\,
{\cal D}(i|j)\bigg]^2\bigg\rangle_{\bar Q}\,.\ee

The remaining average over $\bar Q$ is not so easy to perform, so we shall
evaluate it by using various approximations. To understand the nature of these approximations,
it is useful to consider first the lowest order term in the expansion of the exponential,
which corresponds to the single--scattering approximation (as obvious after comparing
with Eq.~(\ref{Seikonal2})):
\be\label{onescatt}
S_{N\times N'}^{\rm \,one-scatt}&=& 1 -\frac{\lambda}{2} \sum_{i=1}^N
\bigg\langle \sum_{j=1}^{N'}\sum_{m=1}^{N'} \bar Q_j^a \bar Q_m^a\,
{\cal D}(i|j){\cal D}(i|m)\bigg\rangle_{\bar Q}\nn
&=& 1 -\frac{\lambda^2 N_g}{2} \sum_{i=1}^N \sum_{j=1}^{N'} \big[{\cal D}(i|j)\big]^2.\ee
Since $\lambda^2 N_g\approx g^4/4$ at large $N_c$, this 
contribution to the scattering amplitude $T \equiv 1 - S$ is of order
$\alpha_s^2 NN'$. After averaging with the BFKL weight functions, as in
Eq.~(\ref{SY1}), this yields a contribution of order $(\alpha_s N(Y/2))^2\sim
\alpha_s^2 \,{\rm e}^{\omega_0 Y}$ (``the single pomeron exchange''),
 which becomes of order one for $Y\sim Y_c/2$,
cf. Eq.~(\ref{Yc}). Thus, for rapidities $Y\simge Y_c/2$, the
single--scattering approximation breaks down, and some of the higher order terms 
in the expansion of the exponential in Eq.~(\ref{SY2}) must be taken into account.

Specifically, we need to include correctly the terms of order $(\alpha_s^2 NN')^{n}$
for any $n$, since they contribute to leading order, but we can neglect terms with fewer
powers of $N$ or $N'$ (for a given power of $\alpha_s^2$), since, e.g., $\alpha_s^2 N\ll 1$
in the regime of interest, cf. Eq.~(\ref{range}). The strategy to isolate such
terms is similar to that used in the construction of the weight function
in Sect. 3.3. Namely, we first rewrite the exponential in Eq.~(\ref{SY2}) as a product
of $N$ exponentials, one for each dipole, and then we keep only the second order
term in the expansion of each factor in this product:
\be\label{SY3}
S_{N\times N'}\,\approx\,\bigg\langle \prod_{i=1}^N \,\bigg\{1
-\frac{\lambda}{2} \sum_{j=1}^{N'}\sum_{m=1}^{N'} \bar Q_j^a \bar Q_m^a\,
{\cal D}(i|j){\cal D}(i|m)\bigg\}\bigg\rangle_{\bar Q}\,.\ee
This is justified since the exponent which has been expanded out is of order
$\alpha_s^2 N'$ (after averaging over color), and thus is genuinely small.
Physically, this means that for a given dipole in the right--moving onium, 
it is sufficient to consider its {\it single scattering} with
any of the dipoles in the left--moving onium. This is similar to the discussion leading
to Eq.~(\ref{Sexp}) for dipole--hadron scattering: whenever the hadron is characterized
by weak fields (as is the case here for both onia), the scattering of a single
dipole can be computed in the two--gluon exchange approximation. 
We thus see that, in the present context, the multiple scattering
consists in the {\it simultaneous scattering of several pairs of dipoles from the two onia},
while each individual dipole undergoes, at most, single scattering.

But even after the expansion leading to Eq.~(\ref{SY3}), the color average over $\bar Q$
remains difficult to perform, and will be evaluated here only in the large--$N_c$ limit.
This requires some clarifications, since the $N_c$--counting turns to be quite
different in the construction of the wavefunction and in the scattering problem.
Recall that, the dominant effects that have been resummed in the construction 
of the weight function in Sect. 3 were terms of order $(\alpha_s N_c Y)^n$, with $n\ge 1$.
In the present discussion of high--energy scattering, the dominant effects that we are about 
to include are powers of $\alpha_s^2 N N'\sim \alpha_s^2 \,{\rm e}^{\omega_0 Y}$. 
Thus, although their $N_c$ content is very different, the effects of order
$\alpha_s N_c Y$ and $\alpha_s^2 \,{\rm e}^{\omega_0 Y}$ are really treated
on the same footing, namely, as effects of order one. The large--$N_c$ approximation
is then obtained by neglecting terms which are suppressed by negative powers of $N_c$
as compared to these leading order terms.
As we show now, the calculation of the color average in Eq.~(\ref{SY3})
greatly simplifies in this limit.

\begin{figure}
  \centerline{
  \epsfsize=0.8\textwidth
\epsfbox{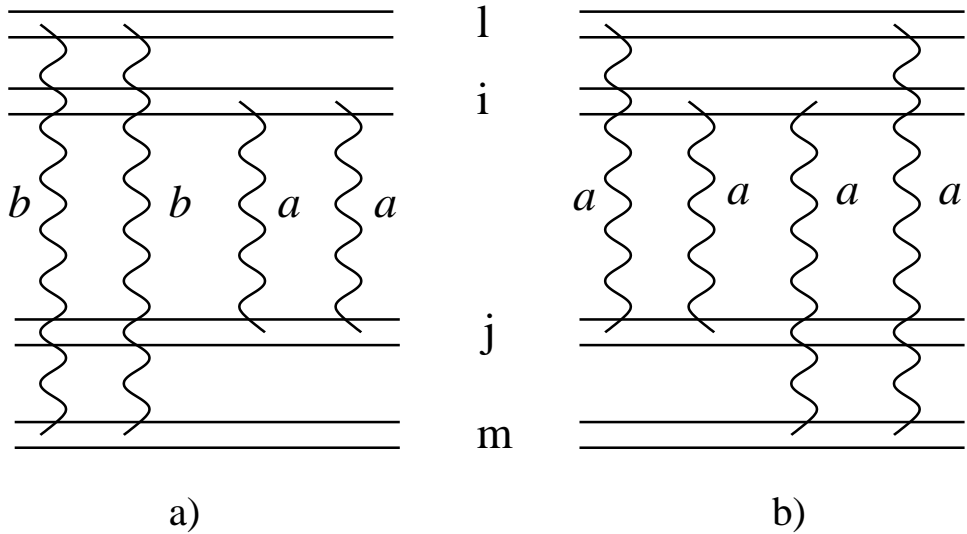}
  }
 \caption[]{The two types of two pomeron exchanges according to Eq.~(\ref{2Pcolor}).
The dipoles $i$ and $l$ are right--movers, while $j$ and $m$ are left--movers.
The label $a$ or $b$ on a gluon line denotes the color exchanged through that line.}
\label{2POMERON}
\end{figure}

Consider, for definiteness, the term quadratic in $\lambda$ in the expansion of the
product in Eq.~(\ref{SY3}). (The term linear in $\lambda$ has been already evaluated in
Eq.~(\ref{onescatt}).) This term describes the simultaneous scattering of two pairs
of dipoles, or  ``two pomeron exchange'':
\be\label{2pom}\hspace*{-.7cm}
T_{N\times N'}^{\rm \,2\,pomeron}=
\bigg(\frac{\lambda}{2}\bigg)^2 \!\sum_{1\le i < l \le N}
\bigg\langle  \sum_{j,m=1}^{N'}\bar Q_j^a \bar Q_m^a\,
{\cal D}(i|j){\cal D}(i|m)
\sum_{j',m'=1}^{N'} \bar Q_{j'}^b \bar Q_{m'}^b\,
{\cal D}(l|j'){\cal D}(l|m')
\bigg\rangle_{\bar Q}.\,\,\ee
Note that there are $N(N-1)/2\times (N')^2$ terms altogether, so the contribution
(\ref{2pom}) is of order $(\alpha_s^2 N N')^2$, as expected. The $N_c$--counting becomes
transparent after averaging over color (cf. Eq.~(\ref{QLR})) :
\be\label{2Pcolor}
\Big\langle \bar Q_j^a \bar Q_m^a\,\bar Q_{j'}^b \bar Q_{m'}^b\Big\rangle_{\bar Q}
\,=\,{\lambda}^2\Big\{N^2_g \delta_{jm}\delta_{j'm'} + N_g\big(\delta_{jj'}\delta_{mm'}
+\delta_{jm'}\delta_{mj'}\big)\Big\}.\ee
The two contributions within the braces --- of order $N^2_g$ and $N_g$, respectively ---
are illustrated in Fig.~\ref{2POMERON}. The contribution in Fig.~\ref{2POMERON}.a,
which is dominant at large $N_c$, involves the independent scattering between two pairs
of dipoles: the pair $(i,j)$ and the pair $(l,m)$. Each collision brings in a factor
$N_g=N_c^2-1$ from the sum over colors. In the process in Fig.~\ref{2POMERON}.b, the sum
over colors gets closed only after mixing all the four dipoles ($l \to j \to i\to m \to l$),
thus giving rise to only one factor of $N_g$. This second process will be neglected in the 
large--$N_c$ approximation.

The generalization to the higher order terms in the expansion of the
product in Eq.~(\ref{SY3}) is  immediate: the large--$N_c$ limit is tantamount
to performing the average over color {\it independently} for each of the $N$ factors
in this product. This  yields:
\be\label{SY4}
S_{N\times N'}\,\approx\, \prod_{i=1}^N \,\bigg\{1
-\frac{\lambda^2 N_g}{2} \sum_{j=1}^{N'}\big[{\cal D}(i|j)\big]^2\bigg\}\,,\ee
which is recognized as the $S$--matrix element for a right--moving system of $N$ dipoles 
which scatter  {\it independently} off the color field created by a left--moving system
of $N'$ dipoles. Note, however, that such an independence holds only for the
scattering between {\it given configurations} of dipoles. After averaging over
all the configurations as in Eq.~(\ref{SY1}), the scatterings get 
correlated with each other, because of the correlations included 
in the weight functions. It is remarkable that the unitarity constraint
$S_{N\times N'} \le 1$ is satisfied already for fixed configurations (and not only in the
average), as obvious on Eq.~(\ref{SY4}). 

To the accuracy of interest,  Eq.~(\ref{SY4}) can be rewritten in the
manifestly symmetric form:
\be\label{SYfinal}
S_{N\times N'}\,=\, {\rm exp}\bigg\{\!
-\frac{\lambda^2 N_g}{2} \sum_{i=1}^N\sum_{j=1}^{N'}\big[{\cal D}(i|j)\big]^2\bigg\}\,,\ee
which is formally like a Glauber approximation: The multiple scattering series is resummed
as the exponential of minus the amplitude (\ref{onescatt}) for a single scattering.
But, once again, this exponentiation holds only {\it configuration by configuration}.
After averaging over all such configurations, the resulting $S$--matrix
for onium--onium scattering:
\be\label{SOO}\hspace*{-.7cm}
S_Y=\sum_{N=1}^\infty\int \! d\Gamma_N\,P_N(Y/2)\,
\sum_{N'=1}^\infty\int \! d\Gamma_{N'}\,P_{N'}(Y/2)\,\,{\rm exp}\bigg\{\!
-\frac{\lambda^2 N_g}{2} \sum_{i=1}^N\sum_{j=1}^{N'}\big[{\cal D}(i|j)\big]^2\bigg\},\,\,\ee
differs significantly from a simple exponential of the ``single pomeron exchange''.
As discussed in Refs. \cite{AM95,AMSalam96,IMfluct}, this difference has dramatical
consequences in the  high--energy regime where the $S$--matrix is very small.
Eq.~(\ref{SOO}) coincides, as  anticipated, 
 with the formula proposed in the framework of the
color dipole picture in Ref. \cite{AM95}, and used for numerical studies of unitarization
in Refs. \cite{Salam95,AMSalam96}.

\vspace*{.5cm}
\section*{Acknowledgments}

We would like to thank Jean--Paul Blaizot, Larry McLerran and Robi Peschanski for 
useful conversations and incentive remarks. Much of this work was done
while one of the authors (A.M.) was a visitor at LPT (Universit\'e de
Paris XI, Orsay). He wishes to thank Dr. Dominique Schiff for her 
hospitality and support during this visit.

\appendix
\section{Appendix}
\setcounter{equation}{0}

In this Appendix we collect some of the technical details that have been omitted
in the discussion of the quantum evolution of the onium weight function in Sect. 3. 
Specifically, we shall verify that the recurrence formula (\ref{recurP}) follows
indeed from the functional evolution in Eq.~(\ref{WYdY}), then we shall check the
probability conservation in Eqs.~(\ref{PROBY})--(\ref{PROBN}), and finally we shall
introduce the dipole number density and  deduce the corresponding evolution equation.
This turns out to be the BFKL equation,  as expected from the corresponding analysis 
within CDP \cite{AM94}.

It turns out that the key technical step behind all the subsequent manipulations is a special 
change of variables within the integral over the transverse coordinates of the dipoles, 
with measure (\ref{GAMMAN}). Let us illustrate this with the derivation of the gain term
in the r.h.s. of Eq.~(\ref{recurP}). (The loss term there can be
trivially inferred from Eq.~(\ref{WYdY}).)

Start with the gain term corresponding to a given value of $N$ in the r.h.s. 
of Eq.~(\ref{WYdY}). This means that we are studying the evolution from a
configuration with $N$ dipoles at rapidity $Y$ to a configuration with $N+1$ dipoles 
at rapidity $Y+{\rm d}Y$. This term is rewritten here for convenience:
\be\label{Wgain}
\bar\alpha_s\,{\rm d}Y\, \int {\rm d}^2\bm{x}_1{\rm d}^2\bm{x}_2\dots{\rm d}^2\bm{x}_{N-1}
{\rm d}^2\bm{z}\,\,P_N(\bm{x}_1,\dots,\bm{x}_{N-1}|Y)\nn
\times\,\,\sum_{i=1}^N \,\bigg(\prod_{j\ne i} [\bm{x}_{j-1},\bm{x}_j]\bigg)\,[\bm{x}_{i-1},\bm{z}]
[\bm{z},\bm{x}_i]\,\,{\cal M}({\bm{x}}_{i-1},{\bm{x}}_i,{\bm z}),
\ee
in compact notations where $\bar\alpha_s \equiv g^2N_c/2\pi$ and, e.g.,
\be
[\bm{x}_{i-1},\bm{z}]\equiv \bigg[1+\frac{\lambda}{2}
\bigg(\int_{\bm x}\,{\cal G}({\bm x}|{\bm{x}}_{i-1},{\bm{z}})
\frac{\delta}{\delta \alpha^a_{\bm{x}}} \bigg)^2\bigg].\ee
(The functional derivatives act on $\delta[\alpha]$, which is omitted here, for simplicity.)
Note that we are using the dipole labelling introduced above Eq.~(\ref{GAMMAN}); thus,
as compared to the original notations in Eq.~(\ref{WYdY}), we have here replaced
$(\bm{x}_i, \bm{y}_i)\longrightarrow ({\bm{x}}_{i-1},\bm{x}_i)$. 

We would like to show that Eq.~(\ref{Wgain}) can be equivalently rewritten as
\be\label{Wgain1}
 \int {\rm d}^2\bm{x}_1{\rm d}^2\bm{x}_2\dots{\rm d}^2\bm{x}_{N}
\,\,{\rm d} P_{N+1}^{(g)}(\bm{x}_1,\dots,\bm{x}_{N}|Y)\,\prod_{i=1}^{N+1}\,
[\bm{x}_{i-1},\bm{x}_i]\,,\ee
where ${\rm d} P_{N+1}^{(g)}$ is the gain term in Eq.~(\ref{recurP}) with $N\to N+1$.
With this aim, take a particular term in the sum over $i$ in Eq.~(\ref{Wgain})
(say, the $i$th term), and change the integration variables as follows:
\be\label{change}
\bm{x}_1 \to \bm{y}_1,\,\dots\,,\,\,{\bm{x}}_{i-1}\to \bm{y}_{i-1}\,,\,\,\,\bm{z}\to \bm{y}_i\,,\,
\,\,\bm{x}_i\to \bm{y}_{i+1}\,,\,\dots,\,\,\,{\bm{x}}_{N-1}\to \bm{y}_{N}.\ee
This implies ${\cal M}({\bm{x}}_{i-1},{\bm{x}}_i,{\bm z}) \to {\cal M}(\bm{y}_{i-1},\bm{y}_{i+1},
\bm{y}_i)$ and $$P_N(\bm{x}_1,\dots,\bm{x}_{N-1}|Y) \to 
P_{N}(\bm{y}_1,\dots,\bm{y}_{i-1},{\bm{y}}_{i+1},\dots,\bm{y}_{N}|Y) \equiv
P_{N}(\bm{y}_1,\dots,\hat{\bm{y}}_i,\dots,\bm{y}_{N}|Y).$$
After this operation, Eq.~(\ref{Wgain}) is rewritten as:
\be
\bar\alpha_s\,{\rm d}Y \int {\rm d}^2\bm{y}_1\dots{\rm d}^2\bm{y}_{N}
\,\prod_{i=1}^{N+1}\,[\bm{y}_{i-1},\bm{y}_i]
\sum_{i=1}^N \, {\cal M}(\bm{y}_{i-1},\bm{y}_{i+1},\bm{y}_i)
P_{N}(\bm{y}_1,\dots,\hat{\bm{y}}_i,\dots,\bm{y}_{N}|Y),\nonumber\ee
which is indeed of the form (\ref{Wgain1}) with (after renoting $\bm{y}_i$ as $\bm{x}_i$) :
\be\label{Wgain2}\hspace*{-.5cm}
{\rm d} P_{N+1}^{(g)}(\bm{x}_1,\dots,\bm{x}_{N}|Y)=
\bar\alpha_s\,{\rm d}Y \sum_{i=1}^N  {\cal M}
({\bm{x}}_{i-1},{\bm{x}}_{i+1},{\bm{x}}_i)
P_{N}(\bm{x}_1,\dots,\hat{\bm{x}}_i,\dots,\bm{x}_{N}|Y)\,\,\,.\ee
As anticipated, this is the gain term in Eq.~(\ref{recurP}) with $N\to N+1$.

To verify the probability conservation in its most stringent form, namely
(cf. Eq.~(\ref{PROBN})),
\be\label{PROBN1}
\int  d\Gamma_N\,\,{\rm d} P_N^{(l)}(Y)\,+\,\int  d\Gamma_{N+1}
\,\,{\rm d} P_{N+1}^{(g)}(Y)\,=\,0,\ee
start with Eq.~(\ref{Wgain2}) for ${\rm d} P_{N+1}^{(g)}(Y)$, and perform
the change of variables in Eq.~(\ref{change}) in reverse order, to deduce:
\be\hspace*{-.7cm}
\int \! d\Gamma_{N+1}
\,{\rm d} P_{N+1}^{(g)}(Y)=\bar\alpha_s\,{\rm d}Y\!\int {\rm d}^2\bm{x}_1\dots{\rm d}^2\bm{x}_{N-1}
\,\,P_N(\bm{x}_1,\dots,\bm{x}_{N-1}|Y) \sum_{i=1}^N 
\int_{\bm{z}}{\cal M}({\bm{x}}_{i-1},{\bm{x}}_i,{\bm z}),\nonumber\ee
which is recognized indeed as $-\int  d\Gamma_N\,\,{\rm d} P_N^{(l)}$, according
to Eq.~(\ref{recurP}). 

Let us finally introduce observables which are obtained by averaging over
the transverse coordinates of the dipoles, and show how to construct evolution
equations for them. An example is the dipole number density, defined as follows:
\be\label{Ddensity}\hspace*{-.7cm}
n_Y(\bm{x},\bm{y}|\bm{x}_0,\bm{y}_0)\,\equiv\,
\sum_{N=1}^\infty\int \! d\Gamma_N\,P_N(\bm{x}_1,\dots,\bm{x}_{N-1}|Y)\,
\sum_{i=1}^N \,
\delta^{(2)}({\bm{x}}_{i-1}-{\bm{x}})\delta^{(2)}({\bm{x}}_i-{\bm{y}}).\,\,\ee
This is the density of dipoles with the quark located at $\bm{x}$ and the antiquark 
at $\bm{y}$ produced after the evolution of an initial dipole $(\bm{x}_0,\bm{y}_0)$ 
through a rapidity interval equal to $Y$. More generally, an observable of this type,
which is not sensitive to the dipole color charges and fields, but only to their
transverse positions, is computed as follows:
\be\label{DOBS}
{\cal O}(Y)\,=\,\sum_{N=1}^\infty\int \! d\Gamma_N\,P_N(\bm{x}_1,\dots,\bm{x}_{N-1}|Y)\,
{\cal O}_N(\bm{x}_1,\dots,\bm{x}_{N-1}),\ee
and  satisfies an evolution equation which is obtained by taking a
derivative w.r.t. $Y$ in the above equation and using Eq.~(\ref{evolP}) for
${\del P_N}/{\del Y}$ :
\be\hspace*{-.7cm}
\frac{\del {\cal O}(Y)}{\del Y}&=&\sum_{N=1}^\infty\int \! d\Gamma_N\,\bigg(
\frac{\del P_N^{(l)}}{\del Y} + \frac{\del P_N^{(g)}}{\del Y}\bigg)
{\cal O}_N(\bm{x}_1,\dots,\bm{x}_{N-1})\\
&=&\sum_{N=1}^\infty\bigg\{\int \! d\Gamma_N\,\frac{\del P_N^{(l)}}{\del Y}\,
{\cal O}_N(\bm{x}_1,\dots,\bm{x}_{N-1})\,+\int  d\Gamma_{N+1}\,
\frac{\del P_{N+1}^{(g)}}{\del Y}\,{\cal O}_{N+1}(\bm{x}_1,\dots,\bm{x}_{N})
\bigg\},\nonumber\ee
where we have used the fact that the gain term vanishes for $N=1$.
Via the same manipulations as before (namely, the change of variables 
in Eq.~(\ref{change}), but in reverse order), one readily obtains:
\be\label{ODY}\hspace*{-.7cm}
\frac{\del {\cal O}(Y)}{\del Y}&= &\bar\alpha_s\sum_{N=1}^\infty
\int d\Gamma_N\,P_N(\bm{x}_1,\dots,\bm{x}_{N-1}|Y)\,
\sum_{i=1}^N \int_{\bm{z}}{\cal M}({\bm{x}}_{i-1},{\bm{x}}_i,{\bm z})\nn
&{}&
\times\,\,\,\,\Big\{- {\cal O}_N(\bm{x}_1,\dots,\bm{x}_{N-1}) + {\cal O}_{N+1}(\bm{x}_1,\dots,
{\bm{x}}_{i-1},{\bm{z}},{\bm{x}}_i,\dots,\bm{x}_{N-1})\Big\}.\ee
If ${\cal O}_N$ is the sum of $N$ terms (one for each dipole),
so like in Eq.~(\ref{Ddensity}), it is clear that the only such terms which survive in
the evolution equation (\ref{ODY}) are those associated with the dipole which 
has split in the course of evolution. All the other terms cancel in between the 
loss and gain contributions. 

Specifically, for the dipole number density (\ref{Ddensity}), Eq.~(\ref{ODY}) gives:
\be\label{nDY}\hspace*{-.7cm}
\frac{\del n_Y}{\del Y}&= &\bar\alpha_s\sum_{N=1}^\infty
\int d\Gamma_N\,P_N(\bm{x}_1,\dots,\bm{x}_{N-1}|Y)\,
\sum_{i=1}^N \int_{\bm{z}}{\cal M}({\bm{x}}_{i-1},{\bm{x}}_i,{\bm z})\\
&{}&\times\,\,\Big\{- 
\delta({\bm{x}}_{i-1}-{\bm{x}})\delta({\bm{x}}_i-{\bm{y}})
+ \delta({\bm{x}}_{i-1}-{\bm{x}})\delta({\bm{z}}-{\bm{y}})
+ \delta({\bm{z}}-{\bm{x}})\delta({\bm{x}}_i-{\bm{y}})
\Big\}.\nonumber\ee
After simple manipulations, this is rewritten as a closed equation for $n_Y$ :
\be\label{nY}\hspace*{-.7cm}
\frac{\del }{\del Y}\,n_Y(\bm{x},\bm{y}|\bm{x}_0,\bm{y}_0)\!\!&=&\!\!\bar\alpha_s \int_{\bm{z}}
\Big\{- {\cal M}({\bm{x}},{\bm{y}},{\bm{z}})n_Y(\bm{x},\bm{y}|\bm{x}_0,\bm{y}_0)
\\
&{}&\,+\,\, {\cal M}({\bm{x}},{\bm{z}},{\bm{y}})n_Y(\bm{x},\bm{z}|\bm{x}_0,\bm{y}_0)
+ {\cal M}({\bm{z}},{\bm{y}},{\bm{x}})n_Y(\bm{z},\bm{y}|\bm{x}_0,\bm{y}_0)\Big\}.\nonumber\ee
This must be solved with the following initial condition (cf. Eq.~(\ref{Ddensity})
with $P_N\to \delta_{N1}$) :
\be\label{nD0}
n_0(\bm{x},\bm{y}|\bm{x}_0,\bm{y}_0)\,=\,
\delta^{(2)}({\bm{x}}_{0}-{\bm{x}})\delta^{(2)}({\bm{y}}_0-{\bm{y}}).\ee
It is easy to see that Eq.~(\ref{nY}) is free of both infrared and ultraviolet
singularities. It is probably less obvious, but nevertheless true, that this 
equation together with the initial condition (\ref{nD0}) is equivalent
to the following equation, which puts the evolution into the {\it original} dipole 
$(\bm{x}_0,\bm{y}_0)$ :
\be\label{nYBFKL}\hspace*{-.7cm}
\frac{\del }{\del Y}\,n_Y(\bm{x},\bm{y}|\bm{x}_0,\bm{y}_0)\!\!&=&\!\!\bar\alpha_s \int_{\bm{z}}
{\cal M}({\bm{x}_0},{\bm{y}_0},{\bm{z}})\Big\{- n_Y(\bm{x},\bm{y}|\bm{x}_0,\bm{y}_0)\nn
&{}&\qquad
\qquad\qquad +\,\, n_Y(\bm{x},\bm{y}|\bm{x}_0,\bm{z}) +\,
n_Y(\bm{x},\bm{y}|\bm{z},\bm{y}_0)\Big\}.\ee
This is recognized as the BFKL equation, and is the  standard way to describe the
dipole number evolution in CDP \cite{AM94,AM95}.
The equivalence between Eqs.~(\ref{nY}) and (\ref{nYBFKL}) is easy to demonstrate once one
realizes that the solution to any of these equations (with the initial condition (\ref{nD0}))
has the following symmetry property, which can be checked
on the explicit solution to Eq.~(\ref{nYBFKL}), as found e.g. in Refs. \cite{AM95,NW97} :
\be
(\bm{x}-\bm{y})^4 \,n_Y(\bm{x},\bm{y}|\bm{x}_0,\bm{y}_0)\,=\,(\bm{x}_0-\bm{y}_0)^4 \,
n_Y(\bm{x}_0,\bm{y}_0|\bm{x},\bm{y})\,.\ee
Eqs.~(\ref{nY}) and (\ref{nYBFKL}) describe the same physical process --- the BFKL
evolution of a system of dipoles ---, but they approach this evolution from different ends.
In Eq.~(\ref{nY}), the evolution proceeds via dipole splitting at the highest rapidity end:
When $Y$ gets increased by d$Y$, it is the measured dipole $(\bm{x},\bm{y})$ which
gets created or annihilated. By contrast, in Eq.~(\ref{nYBFKL}) the increase in rapidity
is used to ``push backwards'' the original dipole $({\bm{x}_0},{\bm{y}_0})$ by an
amount d$Y$. Then, this dipoles undergoes an additional splitting, which appears
as being earlier with respect to the dipole configuration at rapidity $Y$.
Thus, from this perspective, the system grows via splitting at its lowest rapidity end.

Note finally that, by using the definition  (\ref{Ddensity}) for the dipole number density,
the onium--onium scattering amplitude in the single pomeron exchange approximation
(i.e., the first non--trivial term in the expansion of the exponential in Eq.~(\ref{SOO}))
can be written in the familiar form \cite{AM94,AM95,NW97} :
\be\label{1POMY}\hspace*{-.7cm}
T_{Y}^{\rm \,1\,pomeron}(\bm{x}_0,\bm{y}_0|\bm{x}_1,\bm{y}_1)&\!\!=\!\!&
\int {\rm d}^2\bm{x} \,{\rm d}^2\bar{\bm{x}}\,{\rm d}^2{\bm y}\, {\rm d}^2\bar{\bm{y}}\,\,
\,\,n_{Y-y}({\bm{x}},\bar{\bm{x}}|\bm{x}_0,\bm{y}_0)\nn
&{}&\qquad\quad\times\,
\,T^{\rm 2-gluon}({\bm{x}},\bar{\bm{x}}|{\bm y},\bar{\bm{y}})\,\,
\,\,n_y({\bm y},\bar{\bm{y}}|\bm{x}_1,\bm{y}_1),\ee
where $(\bm{x}_0,\bm{y}_0)$ and $(\bm{x}_1,\bm{y}_1)$ are the parent dipoles
in the two onia, and 
$$T^{\rm 2-gluon}({\bm{x}},\bar{\bm{x}}|{\bm y},\bar{\bm{y}})\,\equiv \,
\frac{\lambda^2 N_g}{2}\,
\big[{\cal D}({\bm{x}},\bar{\bm{x}}|{\bm y},\bar{\bm{y}})\big]^2$$
is the dipole--dipole scattering amplitude in the 2-gluon exchange approximation,
cf. Eqs.~(\ref{Seikonal2})--(\ref{calD}). Eq.~(\ref{nYBFKL}) implies that
the amplitude (\ref{1POMY}) satisfies the BFKL equation, as expected.

\end{document}